\documentclass[aps,prb,twocolumn,showpacs,amsmath,amssymb,groupedaddress,floatfix]{revtex4-2}
\usepackage[utf8]{inputenc}
\usepackage{graphicx}
\usepackage{bm}
\usepackage{latexsym}
\usepackage{physics}
\usepackage{amsfonts}
\usepackage{appendix}
\usepackage[colorlinks = true,
            linkcolor = blue,
            urlcolor  = blue,
            citecolor = blue,
            anchorcolor = blue]{hyperref}
\usepackage{scalerel}

\def\scaleint#1{\vcenter{\hbox{\scaleto[3ex]{\displaystyle\int}{#1}}}}

\newcommand{\hgl}[1]{{\color{black}#1}}
\begin{document}
\title{Accessing the ordered phase of correlated Fermi systems: 
\\vertex bosonization and mean-field theory within the functional renormalization group}
\author{Pietro M. Bonetti}
\affiliation{Max Planck Institute for Solid State Research, Heisenbergstrasse 1, D-70569 Stuttgart}
\date{\today}
\begin{abstract}
We present a consistent fusion of functional renormalization group and mean-field theory which explicitly introduces a bosonic field via a Hubbard-Stratonovich transformation at the critical scale, at which the order sets in. We show that a minimal truncation of the flow equations, that neglects order parameter fluctuations, is integrable and fulfills fundamental constraints as the Goldstone theorem and the Ward identity connected with the broken global symmetry. To introduce the bosonic field, we present a technique to factorize the most singular part of the vertex, even when the full dependence on all its arguments is retained. We test our method on the two-dimensional attractive Hubbard model at half-filling and calculate the superfluid gap as well as the Yukawa couplings and residual two fermion interactions in the ordered phase as functions of fermionic Matsubara frequencies. \hgl{Furthermore, we analyze the gap and the condensate fraction for weak and moderate couplings and compare our results with previous functional renormalization group studies, and with quantum Monte Carlo data.} Our formalism constitutes a convenient starting point for the inclusion of order parameter fluctuations by keeping a full, non simplified, dependence on fermionic momenta and/or frequencies. 
\end{abstract}
\pacs{}
\maketitle
\section{Introduction}
When dealing with correlated Fermi systems, one has very frequently to face the breaking of one or more symmetries of the model through the development of some kind of order. Mean-field theory provides a relatively simple but often qualitatively correct description of \textit{ground state} properties in the ordered phase. Remarkably enough, this is not limited to weak coupling calculations but it may survive at strong coupling~\cite{Eagles1969,Leggett1980}. On the other hand, fluctuations of the order parameter play a key role at \textit{finite temperature} $T$~\cite{Nozieres1985} and in low dimensionalities. In particular, they are fundamental in two-dimensional systems, where they prevent \hgl{continuous symmetry breaking} at any $T\neq0$~\cite{Mermin1966,Hohenberg1967}. \hgl{In the specific case of U(1) or SO(2) symmetry groups, fluctuations} are responsible for the formation of the Berezinskii-Kosterlitz-Thouless (BKT) phase, characterized by quasi long-range order~\cite{Berezinskii1971,Kosterlitz1973}. 

The functional renormalization group (fRG) provides a framework to deal with interacting Fermi systems and ordering tendencies~\cite{Metzner2012,Kopietz2010book}. The inclusion of an infrared cutoff in the bare model allows for the treatment of different energy scales $\Lambda$ in a unified approach. In the most typical cases of symmetry breaking, as those associated with the onset of magnetic or superfluid/superconducting orders, at high energies the system is in its symmetric phase, while by decreasing the scale $\Lambda$, the effective two fermion interaction grows until it reaches a divergence at a scale $\Lambda_c$ in one (or more) specific momentum channel~\cite{Zanchi2000,Halboth2000,Honerkamp2001}. \hgl{This divergence, however, can be an artifact of a poor approximation of the flow equations, such as the 1-loop truncation. Indeed, better approximations, as the 2-loop or the multiloop truncation, can significantly reduce the value of $\Lambda_c$, even down to zero~\cite{Freire2008,Eberlein2014,Hille2020}}. In order to continue the flow into the low energy regime, $\Lambda<\Lambda_c$, one has to explicitly introduce an order parameter taking into account spontaneous symmetry breaking.   
Various approaches are possible. One can, for example, decouple the bare interaction via a Hubbard-Stratonovich transformation and run a flow for a mixed boson-fermion system above~\cite{Schuetz2005,Bartosch2009_II,Isidori2010,Streib2013,Lange2015,Lange2017} and below the critical scale. In this way one is able to study fluctuation effects both in the symmetric and in the ordered phases~\cite{Diehl2007,Diehl2007_II,Strack2008,Bartosch2009,Obert2013,Schuetz2006}. Moreover, the two fermion effective interaction generated by the flow can be re-bosonized scale by scale, with a technique called \textit{flowing bosonization}, either decoupling the bare interaction from the beginning~\cite{Baier2004,Floerchinger2008} or keeping it along the flow, reassigning to the bosonic sector only contributions that arise on top of it~\cite{Krahl2009,Friederich2010,Friederich2011}. A different approach to fluctuation effects consists in including below the critical scale $\Lambda_c$ the anomalous terms arising from the breaking of the global symmetry, by keeping only fermionic degrees of freedom~\cite{Salmhofer2004,Gersch2008,Eberlein2013,Eberlein2014_II,Maier2014}. If one is not interested in the effects of bosonic fluctuations, as it could be for ground state calculations, a relatively simple truncation of flow equations can reproduce a mean-field (MF) like solution~\cite{Salmhofer2004,Gersch2005,Wang2014,Yamase2016}.

Concerning the symmetric phase above the critical scale, recent developments have made the fRG a more reliable method for quantitative and/or strong coupling calculations. We refer, in particular, to the development of the multiloop fRG, that has been proven to be equivalent to the parquet approximation~\cite{Kugler2018_I,Kugler2018_II,Tagliavini2019,Hille2020} and the fusion of the fRG with the dynamical mean-field theory (DMFT)~\cite{Georges1996} in the so called DMF\textsuperscript{2}RG scheme~\cite{Taranto2014,Vilardi2019}. Within these frameworks, the full dependence of the effective two fermion interaction on all three Matsubara frequencies is often kept.

On the other hand, many efforts have been made in order to reduce the computational complexity of the effective interaction with a full dependence on its fermionic arguments. This is mainly achieved by describing the fermion-fermion interaction process through the exchange of a small number of bosons. Many works treat this aspect not only within the fRG~\cite{Friederich2010,Friederich2011,Denz2020}, but also within the DMFT, in the recently introduced single boson exchange approximation~\cite{Krien2019_I,Krien2019_II}, its nonlocal extensions, the TRILEX approach for example~\cite{Ayral2016}, or the dual boson theory~\cite{Rubtsov2012,Stepanov2018,Stepanov2019,Peters2019}. Describing the fermionic interactions in terms of exchanged bosons is important not only to reduce the computational complexity, but also to identify those collective fluctuations that play a fundamental role in the ordered phase.

In this paper, we present a truncation of the fRG flow equations, in which a bosonic field is explicitly introduced, and we prove it to be equivalent to the fusion of the fRG with MF theory introduced in Refs.~\cite{Wang2014,Yamase2016}. These flow equations fulfill fundamental constraints as the Goldstone theorem and the global Ward identity connected with spontaneous symmetry breaking (SSB), and they can be integrated, simplifying the calculation of correlation functions in the ordered phase to a couple of self consistent equations, one for the bosonic field expectation value, and another one for the Yukawa coupling between a fermion and the Goldstone mode. In order to perform the Hubbard-Stratonovich transformation, we decompose the effective two fermion interaction in terms of an exchanged boson, which becomes massless at the critical scale, and a residual interaction, and we present a technique to factorize the fRG vertex when its full dependence on fermionic Matsubara frequencies is kept. We prove the feasibility and efficiency of our formalism by applying it to the two-dimensional half-filled attractive Hubbard model, calculating the superfluid gap, Yukawa couplings and residual two fermion interactions in the SSB phase \hgl{and comparing our results with previous fRG and quantum Monte Carlo studies.} 
One notable aspect of our formalism is that the full dependence on fermionic momenta and/or frequencies can be retained. This makes it suitable for a combination with the newly developed methods within the fRG, to continue the flow with a simple truncation in those cases in which the effective two fermion interaction diverges. In the one loop truncation, both in plain fRG~\cite{Vilardi2017} and in the DMF\textsuperscript{2}RG~\cite{Vilardi2019}, these divergences are actually found at finite temperature, indicating the onset of spontaneous symmetry breaking. 
Our method can be also combined with the multiloop fRG, where no divergences are found at finite temperature in 2D~\cite{Hille2020}, to study three-dimensional systems or zero temperature phases. 
Furthermore, the introduction of the bosonic field makes our method a convenient starting point for the inclusion of order parameter fluctuations on top of the MF, and paves the way for the study of the SSB phases with a full treatment of fermionic Matsubara frequency dependencies.

This paper is organized as follows. In Sec.~\ref{sec: fRG} we give a short overview of the fRG and its application to correlated Fermi systems. In Sec.~\ref{sec: model} we introduce the attractive Hubbard model, that will be the prototypical model for the application of our method. In Sec.~\ref{sect: fermionic formalism} we review the MF approximation within the fRG by making use only of fermionic degrees of freedom. In Sec.~\ref{sect: bosonic formalism} we introduce our method by reformulating the fermionic MF approach with the introduction of a bosonic field and we prove the equivalence of the two methods. In Sec.~\ref{sec: vertex bosonization} we expose a strategy to extract a factorizable part from the effective two fermion interactions, necessary to implement the Hubbard-Stratonovich transformation. 
This strategy is suitable for the application to the most frequently used schemes within the fRG. 
In Sec.~\ref{sec: results} we present some exemplary results for the attractive Hubbard model. A conclusion in Sec.~\ref{sec: conclusion} closes the presentation.
\section{Functional renormalization group}
\label{sec: fRG}
In this section we present a short review of the fRG applied to interacting Fermi systems and we refer to Ref.~\cite{Metzner2012} for further details. 
Providing the bare fermionic action with a regulator $R^\Lambda$, 
\begin{equation}
    \mathcal{S}\left[\psi,\overline{\psi}\right]\rightarrow  \mathcal{S}\left[\psi,\overline{\psi}\right] + \left(\overline{\psi}R^\Lambda,\psi\right),
\end{equation}
where the symbol $(\cdot,\cdot)$ indicates a sum over quantum numbers and fermionic Matsubara frequencies $\nu=(2j+1)\pi T$, with $j\in \mathbb{Z}$, one can derive an exact differential equation for the effective action as a function of the scale $\Lambda$~\cite{Wetterich1993,Berges2002}:
\begin{equation}
    \partial_\Lambda\Gamma^\Lambda[\psi,\overline{\psi}]=-\frac{1}{2}\widetilde{\partial}_\Lambda\text{tr}\ln\left[\mathbf{\Gamma}^{(2)\Lambda}[\psi,\overline{\psi}]+R^\Lambda\right],
    \label{eq: Wetterich eq. ferm}
\end{equation}
where $\mathbf{\Gamma}^{(2)\Lambda}$ is the matrix of second derivatives of the effective action w.r.t. the fermionic fields, $\widetilde{\partial}_\Lambda$ is a derivative acting only on the explicit $\Lambda$-dependence of $R^\Lambda$ and the trace is intended to run over all the quantum numbers and Matsubara frequencies. In general, the regulator can be any generic function of the scale $\Lambda$ and the fermionic "$d$+1 momentum" $k=(\mathbf{k},\nu)$ (with $\mathbf{k}$ being the spatial momentum), provided that $R^{\Lambda\rightarrow\Lambda_\text{init}}\rightarrow \infty$ and $R^{\Lambda\rightarrow\Lambda_\text{fin}}\rightarrow 0$. In this way, Eq.~\eqref{eq: Wetterich eq. ferm} can be complemented with the initial condition
\begin{equation}
    \Gamma^{\Lambda=\Lambda_\text{init}}[\psi,\overline{\psi}]=\mathcal{S}[\psi,\overline{\psi}].
\end{equation}
Eq.~\eqref{eq: Wetterich eq. ferm} is however very hard to tackle. A common procedure is to expand the effective action $\Gamma^\Lambda$ in polynomials of the fields up to a finite order, so that one is limited to work with a finite number of scale dependent couplings. Rather often, in the context of correlated Fermi systems, this truncation is restricted to a flow equation for the self-energy $\Sigma^\Lambda$ and a vertex $V^\Lambda$, describing the two fermion effective interaction. The differential equations for these couplings can be inferred directly from Eq.~\eqref{eq: Wetterich eq. ferm}. Furthermore, when working with systems that possess U(1) charge, SU(2) spin rotation and translational symmetries, the vertex $V^\Lambda$ as a function of the spin variables $\sigma_i$ and the four $d+1$ momenta $k_i$ of the fermions (two incoming, two outgoing) can be written as
\begin{equation}
    \begin{split}
        &V^\Lambda_{\sigma_1\sigma_2\sigma_3\sigma_4}(k_1,k_2,k_3)=\\
        &V^\Lambda(k_1,k_2,k_3)\delta_{\sigma_1\sigma_4}\delta_{\sigma_2\sigma_3}-V^\Lambda(k_2,k_1,k_3)\delta_{\sigma_1\sigma_3}\delta_{\sigma_2\sigma_4},
    \end{split}
\end{equation}
where the fermions labeled as 1-2 are considered as incoming and the ones labeled as 3-4 as outgoing in the scattering process. Furthermore, thanks to translational invariance, the vertex is nonzero only when the total momentum is conserved, that is when $k_1+k_2=k_3+k_4$. So, one can shorten the momentum dependence to three momenta, the fourth being fixed by the conservation law. By exploiting the relation above, one is left with the calculation of a single coupling function $V^\Lambda$ that summarizes all possible spin combinations. Its flow equation reads (dropping momentum dependencies for the sake of compactness)
\begin{equation}
    \partial_\Lambda V^\Lambda = \mathcal{T}_\text{pp}^\Lambda+\mathcal{T}_\text{ph}^\Lambda+\mathcal{T}_\text{phx}^\Lambda+\Gamma^{(6)\Lambda} \circ \widetilde{\partial}_\Lambda G^\Lambda,
    \label{eq: vertex flow equation symm}
\end{equation}
where the last term contains the 3-fermion coupling $\Gamma^{(6)\Lambda}$ contracted with the \textit{single scale propagator} $\widetilde{\partial}_\Lambda G^\Lambda$. This term is often neglected or treated in an approximate fashion in most applications. The remaining three terms can be expressed as loop integrals involving two fermionic propagators and two vertices $V^\Lambda$. They are grouped in three channels, namely \textit{particle-particle} ($ \mathcal{T}_\text{pp}^\Lambda$), \textit{particle-hole} ($ \mathcal{T}_\text{ph}^\Lambda$) and \textit{particle-hole-crossed} ($ \mathcal{T}_\text{phx}^\Lambda$), depending on which combination of momenta is transported by the loop. For the expressions of all the terms in Eq.~\eqref{eq: vertex flow equation symm} see Ref.~\cite{Metzner2012}.

In numerous applications of the fRG to various systems, the vertex function $V^\Lambda$ diverges before the numerical integration of Eq.~\eqref{eq: vertex flow equation symm} reaches the final scale $\Lambda_\text{fin}$. This fact signals the tendency of the system to develop some kind of order by spontaneously breaking one (or more) of its symmetries. One can often trace back the nature of the order tendency by looking at which of the terms in Eq.~\eqref{eq: vertex flow equation symm} contributes the most to the flow of $V^\Lambda$ near the critical scale $\Lambda_c$, where the divergence occurs. 
\section{Model}
\label{sec: model}
In this section we present the prototypical model that we use for the application of our method. This is the two-dimensional (2D) attractive Hubbard model, that exhibits an instability in the particle-particle channel, signaling the onset of spin-singlet superfluidity. Our formalism, however, can be extended to a wide class of models, including the 2D repulsive Hubbard model, to study the phases in which (generally incommensurate) antiferromagnetism and/or d-wave superconductivity appear.  
The bare action of the model describes spin-$\frac{1}{2}$ fermions on a 2D lattice experiencing an attractive on-site attraction
\begin{equation}
    \begin{split}
        \mathcal{S}=&-\int_{k,\sigma} \overline{\psi}_{k,\sigma} \left[i\nu-\xi_\mathbf{k}\right]\psi_{k,\sigma} \\
        &+ U \int_{k,k',q} \overline{\psi}_{k,\uparrow} \overline{\psi}_{q-k,\downarrow} \psi_{q-k',\downarrow} \, \psi_{k',\uparrow},
    \end{split}
    \label{eq: bare Hubbard action}
\end{equation}
where $\nu$ is a fermionic Matsubara frequency, $\xi_\mathbf{k}$ is the bare band dispersion measured relative to the chemical potential $\mu$, and $U<0$ is the local interaction. The symbol $\int_k=T\sum_\nu\int\frac{d^2\mathbf{k}}{(2\pi)^2}$ ($T$ being the temperature) denotes an integral over the Brillouin zone and a sum over Matsubara frequencies. 

\hgl{This model, in $d=2$ or 3, at zero or finite temperature, has been subject of extensive studies with several methods, in particular the fRG~\cite{Eberlein2013,Obert2013}, quantum Monte Carlo~\cite{Randeria1992,dosSantos1994,Trivedi1995,Singer1996,Karakuzu2018}, and DMFT and extensions~\cite{Keller2001,Capone2002,Toschi2005,DelRe2019}. }

In the next sections, we will assume that a fRG flow is run for this model, up to a stopping scale $\Lambda_s$, very close to a critical scale $\Lambda_c$ where the vertex $V^\Lambda$ diverges due to a pairing tendency, but still in the symmetric regime. From now on, we will also assume an infrared regulator such that the scale $\Lambda$ is lowered from $\Lambda_\text{ini}$ to $\Lambda_\text{fin}$, so that the inequality $\Lambda_\text{ini}>\Lambda_s\gtrsim\Lambda_c>\Lambda_\text{fin}$ holds. 
\section{Broken symmetry phase: fermionic formalism}
\label{sect: fermionic formalism}
In this section we will present a simple truncation of flow equations that allows to continue the flow beyond $\Lambda_s$ in the superfluid phase within a MF-like approximation, that neglects any kind of order parameter (thermal or quantum) fluctuations. This approximation can be formulated by working only with the physical fermionic degrees of freedom. \\
In order to tackle the breaking of the global U(1) symmetry, we introduce the Nambu spinors
\begin{equation}
    \Psi_k=\left(
    \begin{array}{c}
        \psi_{k,\uparrow}  \\
        \overline{\psi}_{-k,\downarrow}  
    \end{array}
    \right) \hskip 1cm 
    \overline{\Psi}_k=\left(
    \begin{array}{c}
        \overline{\psi}_{k,\uparrow}  \\
        \psi_{-k,\downarrow}  
    \end{array}
    \right).
    \label{eq: Nambu spinors}
\end{equation}
\subsection{Flow equations and integration}
In the SSB phase, the vertex function $V$ acquires anomalous components due to the violation of particle number conservation. In particular, besides the normal vertex describing scattering processes with two incoming and two outgoing particles ($V_{2+2}$), in the superfluid phase also components with three ($V_{3+1}$) or four ($V_{4+0}$) incoming or outgoing particles can arise. We avoid to treat the 3+1 components, since they are related to the coupling of the order parameter to charge fluctuations~\cite{Eberlein2013}, which do not play any role in a MF-like approximation for the superfluid state. It turns out to be useful to work with combinations 
\begin{equation}
    \begin{split}
        &V_\mathcal{A}=\Re\left\{V_{2+2}+V_{4+0}\right\}\\
        &V_\Phi=\Re\left\{V_{2+2}-V_{4+0}\right\},
        \label{eq:  A and Phi vertex combination}
    \end{split}
\end{equation}
that represent two fermion interactions in the longitudinal and transverse order parameter channels, respectively. They are related to the amplitude and phase fluctuations of the superfluid order parameter, respectively. In principle, a longitudinal-transverse mixed interaction can also appear, from the imaginary parts of the vertices in Eq.~\eqref{eq:  A and Phi vertex combination}, but it has no effect in the present MF approximation because it vanishes at zero center of mass frequency~\cite{Eberlein_Thesis}.

Below the stopping scale, $\Lambda <\Lambda_s$, we consider a truncation of the effective action of the form 
\begin{equation}
    \begin{split}
        \Gamma^{\Lambda}_{\text{SSB}}[\Psi,\overline{\Psi}]=-&\int_{k} \overline{\Psi}_{k} \, \left[\mathbf{G}^{\Lambda}(k)\right]^{-1} \Psi_{k}\\
        +&\int_{k,k',q}V^{\Lambda}_{\mathcal{A}}(k,k';q)\, S^1_{k,q}\,S^1_{k',-q}\\
        +&\int_{k,k',q}V^{\Lambda}_{\Phi}(k,k';q)\, S^2_{k,q}\,S^2_{k',-q} ,
    \end{split}
    \label{eq: fermionic SSB truncation}
\end{equation}
with the Nambu bilinears defined as
\begin{equation}
    S^\alpha_{k,q}=\overline{\Psi}_{k}\, \tau^\alpha \,\Psi_{k-q},
    \label{eq: fermion bilinear}
\end{equation}
where the Pauli matrices $\tau^\alpha$ are contracted with Nambu spinor indexes. The fermionic propagator $\mathbf{G}^\Lambda(k)$ is given by the matrix
\begin{equation}
    \left(
    \begin{array}{cc}
        Q_{0}^\Lambda(k)-\Sigma^\Lambda(k) &  \Delta^\Lambda(k)\\
        \Delta^\Lambda(k) & -Q_0^\Lambda(-k)+\Sigma^\Lambda(-k)
    \end{array}
    \right)^{-1},
\end{equation}
where $Q_{0}^\Lambda(k)=i\nu-\xi_\mathbf{k}+R^\Lambda(k)$, $R^\Lambda(k)$ is the regulator, $\Sigma^\Lambda(k)$ is the normal self energy and $\Delta^\Lambda(k)$ is the superfluid gap. The initial conditions at the scale $\Lambda=\Lambda_s$ require $\Delta^{\Lambda_s}$ to be zero and both $V^{\Lambda_s}_\mathcal{A}$ and $V^{\Lambda_s}_\Phi$ to equal the vertex $V^{\Lambda_s}$ in the symmetric phase.

We are now going to introduce the MF approximation to the symmetry broken state, that means that we focus on the $q=0$ component of $V_\mathcal{A}$ and $V_\Phi$ and neglect all the rest. So, from now on we drop all the $q$-dependencies. We neglect the flow of the normal self-energy below $\Lambda_s$, that would require the inclusion of charge fluctuations in the SSB phase, which is beyond the MF approximation. In order to simplify the presentation, we introduce a matrix-vector notation for the gaps and vertices. In particular, the functions $V_\mathcal{A}$ and $V_\Phi$ are matrices in the indices $k$ and $k'$, while the gap and the fermionic propagator behave as vectors. For example, in this notation an object of the type $\int_{k'}V_\mathcal{A}^\Lambda(k,k')\Delta^\Lambda(k')$ can be viewed as a matrix-vector product, $V_\mathcal{A}^\Lambda \Delta^\Lambda$. 

Within our MF approximation, we consider in our set of flow equations only the terms that involve only the $q=0$ components of the functions $V_\mathcal{A}$ and $V_\Phi$. This means that in a generalization of Eq.~\eqref{eq: vertex flow equation symm} to the SSB phase, we consider only the particle-particle contributions. In formulas we have:
\begin{align}
    &\partial_\Lambda V_\mathcal{A}^\Lambda=V_\mathcal{A}^\Lambda\left[\widetilde{\partial}_\Lambda\Pi^\Lambda_{11}\right] V_\mathcal{A}^\Lambda+\Gamma^{(6)\Lambda} \circ \widetilde{\partial}_\Lambda G^\Lambda,
    \label{eq: flow eq Va fermionic}\\
    &\partial_\Lambda V_\Phi^\Lambda=V_\Phi^\Lambda \left[\widetilde{\partial}_\Lambda\Pi^\Lambda_{22}\right] V_\Phi^\Lambda+\Gamma^{(6)\Lambda} \circ \widetilde{\partial}_\Lambda G^\Lambda,
    \label{eq: flow eq Vphi fermionic}
\end{align}
where we have defined the bubbles 
\begin{equation}
    \Pi^\Lambda_{\alpha\beta}(k,k')=-\frac{1}{2}\Tr\left[\tau^\alpha\,\mathbf{G}^\Lambda(k)\,\tau^\beta\,\mathbf{G}^\Lambda(k)\right]\delta_{k,k'},
\end{equation}
where $\delta_{k,k'}=(2\pi)^2/T \,\delta^{(2)}(\mathbf{k}-\mathbf{k}')\delta_{\nu\nu'}$, and the trace runs over Nambu spin indexes.
The last terms of Eqs.~\eqref{eq: flow eq Va fermionic} and~\eqref{eq: flow eq Vphi fermionic} involve the 6-particle interaction, which we treat here in the Katanin approximation, that allows us to replace the derivative acting on the regulator $\widetilde{\partial}_\Lambda$ of the bubbles with the full scale derivative $\partial_\Lambda$~\cite{Katanin2004}. This approach is useful for it provides the exact solution of mean-field models, such as the reduced BCS, in which the bare interaction is restricted to the zero center of mass momentum channel~\cite{Salmhofer2004}. 
In this way, the flow equation~\eqref{eq: flow eq Va fermionic} for the vertex $V_\mathcal{A}$, together with the initial condition $V_\mathcal{A}^{\Lambda_s}=V^{\Lambda_s}$ can be integrated analytically, giving
\begin{equation}
    \begin{split}
        V_\mathcal{A}^\Lambda = &\left[1+V^{\Lambda_s}(\Pi^{\Lambda_s}-\Pi_{11}^\Lambda)\right]^{-1}V^{\Lambda_s}\\ =&\left[1-\widetilde{V}^{\Lambda_s}\Pi_{11}^\Lambda\right]^{-1}\widetilde{V}^{\Lambda_s},
    \end{split}
    \label{eq: Va solution fermionic}
\end{equation}
where  
\begin{equation}
    \Pi^{\Lambda_s}(k,k')=G^{\Lambda_s}(k)G^{\Lambda_s}(-k)\delta_{k,k'},
    \label{eq: bubble at Lambda_s}
\end{equation}
is the (normal) particle-particle bubble at zero center of mass momentum, 
\begin{equation}
    G^{\Lambda}(k)=\frac{1}{Q_0^{\Lambda}(k)-\Sigma^{\Lambda_s}(k)},
    \label{eq: G at Lambda_s}
\end{equation}

is the fermionic normal propagator, and
\begin{equation}
    \widetilde{V}^{\Lambda_s}=\left[1+V^{\Lambda_s}\Pi^{\Lambda_s}\right]^{-1}V^{\Lambda_s}
    \label{eq: irr vertex fermionic}
\end{equation}
is the irreducible (normal) vertex in the particle-particle channel at the stopping scale. The flow equation for the transverse vertex $V_\Phi$ exhibits a formal solution similar to the one in Eq.~\eqref{eq: Va solution fermionic}, but the matrix inside the square brackets is not invertible. We will come to this aspect later. 
\subsection{Gap equation}
Similarly to the flow equations for vertices, in the flow equation of the superfluid gap we neglect the contributions involving the vertices at $q\neq 0$. We are then left with
\begin{equation}
    \partial_\Lambda\Delta^\Lambda(k)=\int_{k'}V_\mathcal{A}^\Lambda(k,k')\,\widetilde{\partial}_\Lambda F^\Lambda(k'),
    \label{eq: gap flow equation}
\end{equation}
where  
\begin{equation}
    F^\Lambda(k)=\frac{\Delta^\Lambda(k)}{[G^\Lambda(k)\,G^\Lambda(-k)]^{-1}+\left[\Delta^\Lambda(k)\right]^2}
    \label{eq: F definition}
\end{equation}
is the anomalous fermionic propagator, with $G$ defined as in Eq.~\eqref{eq: G at Lambda_s}, and with the normal self-energy kept fixed at its value at the stopping scale. By inserting Eq.~\eqref{eq: Va solution fermionic} into Eq.~\eqref{eq: gap flow equation} and using the initial condition $\Delta^{\Lambda_s}=0$, we can analytically integrate the latter, obtaining the gap equation~\cite{Wang2014}
\begin{equation}
    \Delta^\Lambda(k)=\int_{k'}\widetilde{V}^{\Lambda_s}(k,k')\, F^\Lambda(k').
    \label{eq: gap equation fermionic}
\end{equation}
In the particular case in which the contributions to the vertex flow equation from other channels (different from the particle-particle) as well as the 3-fermion interaction and the normal self-energy are neglected also above the stopping scale, the irreducible vertex is nothing but $-U$, the (sign reversed) bare interaction, and Eq.~\eqref{eq: gap equation fermionic} reduces to the standard Hartree-Fock approximation to the SSB state. 
\subsection{Goldstone Theorem}
In this subsection we prove that the present truncation of flow equations fulfills the Goldstone theorem. We revert our attention on the transverse vertex $V_\Phi$. Its flow equation in Eq.~\eqref{eq: flow eq Vphi fermionic} can be (formally) integrated too, together with the initial condition $V_\Phi^{\Lambda_s}=V^{\Lambda_s}$, giving 
\begin{equation}
    \begin{split}
        V_\Phi^\Lambda = &\left[1+V^{\Lambda_s}(\Pi^{\Lambda_s}-\Pi_{22}^\Lambda)\right]^{-1}V^{\Lambda_s}\\ =&\left[1-\widetilde{V}^{\Lambda_s}\Pi_{22}^\Lambda\right]^{-1}\widetilde{V}^{\Lambda_s}.
    \end{split}
    \label{eq: Vphi solution fermionic}
\end{equation}
However, by using the relation 
\begin{equation}
    \Pi_{22}^\Lambda(k,k')=\frac{F^\Lambda(k)}{\Delta^\Lambda(k)}\,\delta_{k,k'},
    \label{eq: Pi22=F/delta}
\end{equation}
one can rewrite the matrix in angular brackets in the second line of Eq.~\eqref{eq: Vphi solution fermionic} as
\begin{equation}
    \delta_{k,k'}-\widetilde{V}^{\Lambda_s}(k,k')\,\frac{F^\Lambda(k')}{\Delta^\Lambda(k')}.
\end{equation}
Multiplying this expression by $\Delta^\Lambda(k')$ and integrating over $k'$, we see that it vanishes if the gap equation~\eqref{eq: gap equation fermionic} is obeyed. Thus, the matrix in angular brackets in Eq.~\eqref{eq: Vphi solution fermionic} has a zero eigenvalue with the superfluid gap as eigenvector. In matrix notation this fact can be expressed as
\begin{equation}
    \left[ 1 - \widetilde{V}^{\Lambda_s}\Pi^\Lambda_{22}\right]\Delta^\Lambda=0.
\end{equation}
Due to the presence of this zero eigenvalue, the above matrix is not invertible. This is nothing but a manifestation of the Goldstone theorem. Indeed, due to the breaking of the global U(1) symmetry, transverse fluctuations of the order parameter become massless at $q=0$, leading to the divergence of the transverse two fermion interaction $V_\Phi$.
\section{Broken symmetry phase: bosonic formalism}
\label{sect: bosonic formalism}
The SSB phase can be accessed also via the introduction of a bosonic field, describing the fluctuations of the order parameter, and whose finite expectation value is related to the formation of anomalous components in the fermionic propagator. In order to introduce this bosonic field, we express the vertex at the stopping scale in the following form:
\begin{equation}
    V^{\Lambda_s}(k,k';q)=\frac{h^{\Lambda_s}(k;q)\,h^{\Lambda_s}(k';q)}{m^{\Lambda_s}(q)}+\mathcal{Q}^{\Lambda_s}(k,k';q).
    \label{eq: vertex at Lambda crit}
\end{equation}
We assume from now on that the divergence of the vertex, due to the appearance of a massless mode, is absorbed into the first term, while the second one remains finite.  In other words, we assume that as the stopping scale $\Lambda_s$ approaches the critical scale $\Lambda_c$ at which the vertex is formally divergent, the (inverse) bosonic propagator $m^{\Lambda_s}(q)$ at zero frequency and momentum vanishes, while the the Yukawa coupling $h^{\Lambda_s}(k;q)$ and the residual two fermion interaction $\mathcal{Q}^{\Lambda_s}(k,k';q)$ remain finite. 

In Sec.~\ref{sec: vertex bosonization} we will introduce a systematic scheme to extract the decomposition~\eqref{eq: vertex at Lambda crit} from a given vertex at the stopping scale.
\subsection{Hubbard-Stratonovich transformation and truncation}
\hgl{Since the effective action at a given scale $\Lambda$ can be viewed as a bare action with bare propagator $G_0-G_0^\Lambda$ (with $G_0^\Lambda$ the regularized bare propagator)~\cite{note_HS_gamma}, one can decouple the factorized (and singular) part of the vertex at $\Lambda_s$ via a Gaussian integration, thus introducing a bosonic field. By adding source terms which couple linearly to this field and to the fermionic ones, one obtains the generating functional of connected Green's functions, whose Legendre transform reads, at the stopping scale}
\begin{equation}
    \begin{split}
        &\Gamma^{\Lambda_s}[\psi,\overline{\psi},\phi]=
        \int_{k,\sigma} \overline{\psi}_{k,\sigma} \left[G^{\Lambda_s}(k)\right]^{-1} \psi_{k,\sigma}\\
        &+\int_{q} \phi^*_q \, m^{\Lambda_s}(q)\, \phi_q\\
        &+\int_{k,k',q}\mathcal{Q}^{\Lambda_s}(k,k';q)\,\overline{\psi}_{k,\uparrow}  \overline{\psi}_{q-k,\downarrow} \psi_{q-k',\downarrow} \psi_{k',\uparrow}\\
        &+\int_{k,q}h^{\Lambda_s}(k;q)\left[ \overline{\psi}_{k,\uparrow} \overline{\psi}_{q-k,\downarrow} \phi_q + \text{h.c.}\right],
    \end{split}
    \label{eq: gamma lambda crit bos}
\end{equation}
\hgl{where $\phi$ represents the expectation value (in presence of sources) of the Hubbard-Stratonovich field.}
Note that we have avoided to introduce an interaction between equal spin fermions. Indeed, since we are focusing on a spin singlet superconducting order parameter, within the MF approximation this interaction has no contribution to the flow equations. 

\hgl{The Hubbard-Stratonovich transformation introduced in Eq.\eqref{eq: gamma lambda crit bos} is free of the so-called Fierz ambiguity, according to which different ways of decoupling of the bare interaction can lead to different mean-field results for the gap (see, for example, Ref.~\cite{Baier2004}).  Indeed, through the inclusion of the residual two fermion interaction, we are able to recover the same equations that one would get without bosonizing the interactions, as proven in Sec.~\ref{subsec: equivalence bos and fer}. In essence, the only ambiguity lies in selecting what to assign to the bosonized part of the vertex and what to $\mathcal{Q}$, but by keeping both of them all along the flow, the results will not depend on this choice.}

We introduce Nambu spinors as in Eq.~\eqref{eq: Nambu spinors} and we decompose the bosonic field into its (flowing) expectation value plus longitudinal ($\sigma$) and transverse ($\pi$) fluctuations~\cite{Obert2013}:
\begin{equation}
    \begin{split}
        &\phi_q=\alpha^\Lambda\,\delta_{q,0} + \sigma_q + i\, \pi_q \\
        &\phi^*_q=\alpha^\Lambda\,\delta_{q,0} + \sigma_{-q} - i\, \pi_{-q},
    \end{split}
\end{equation}
where we have chosen $\alpha^\Lambda$ to be real. For the effective action at $\Lambda<\Lambda_s$ in the SSB phase, we use the following \textit{ansatz}
\begin{equation}
    \begin{split}
        \Gamma^{\Lambda}_\text{SSB}[\Psi,\overline{\Psi},\sigma,\pi]&=\Gamma^\Lambda_{\Psi^2}+\Gamma^\Lambda_{\sigma^2}+\Gamma^\Lambda_{\pi^2}\\
        &+\Gamma^\Lambda_{\Psi^2\sigma} + \Gamma^\Lambda_{\Psi^2\pi}
        +\Gamma^\Lambda_{\Psi^4},
    \end{split}
    \label{eq: bosonic eff action}
\end{equation}
where the first three quadratic terms are given by
\begin{equation}
    \begin{split}
        &\Gamma^\Lambda_{\Psi^2}=-\int_{k} \overline{\Psi}_{k} \left[\mathbf{G}^{\Lambda}(k)\right]^{-1} \Psi_{k}\\
        &\Gamma^\Lambda_{\sigma^2}=-\frac{1}{2}\int_q \sigma_{-q}\,m_\sigma^{\Lambda}(q)\, \sigma_q\\
        &\Gamma^\Lambda_{\pi^2}=-\frac{1}{2}\int_q \pi_{-q}\,m_\pi^{\Lambda}(q)\, \pi_q,    
    \end{split}
\end{equation}
and the fermion-boson interactions are 
\begin{equation}
    \begin{split}
        &\Gamma^\Lambda_{\Psi^2\sigma}=\int_{k,q}h^{\Lambda}_\sigma(k;q)\left\{S^1_{k,-q}\,\sigma_q+ \text{h.c.} \right\}\\
        &\Gamma^\Lambda_{\Psi^2\pi}=\int_{k,q}h^{\Lambda}_\pi(k;q)\left\{S^2_{k,-q}\,\pi_q+ \text{h.c.} \right\},
    \end{split}
\end{equation}
with $S^\alpha_{k,q}$ as in Eq.~\eqref{eq: fermion bilinear}.
The residual two fermion interaction term is written as
\begin{equation}
    \begin{split}
        \Gamma^\Lambda_{\Psi^{4}}=&
        \int_{k,k',q}\mathcal{A}^{\Lambda}(k,k';q)\,S^1_{k,q}\,S^1_{k',-q}\\
        &+\int_{k,k',q}\hskip - 5mm\Phi^{\Lambda}(k,k';q) \,S^2_{k,q}\,S^2_{k',-q}.
    \end{split}    
\end{equation}
As in the fermionic formalism, in the truncation in Eq.~\eqref{eq: bosonic eff action} we have neglected any type of longitudinal-transverse fluctuation mixing in the Yukawa couplings, bosonic propagators and two fermion interactions because at $q=0$ they are identically zero. In the bosonic formulation, as well as for the fermionic one, the MF approximation requires to focus on the $q=0$ components of the various terms appearing in the effective action and neglect all the rest. So, from now on we drop all the $q$-dependencies. We will make use of the matrix notation introduced in Sec.~\ref{sect: fermionic formalism}, for which the newly introduced Yukawa couplings behave as vectors and bosonic inverse propagators as scalars. 
\subsection{Flow equations and integration}
Here we focus on the flow equations for two fermion interactions, Yukawa couplings and bosonic inverse propagators in the longitudinal and transverse channels within a MF approximation, that is we focus only on the Cooper channel ($q=0$) and neglect all the diagrams containing internal bosonic lines or the couplings $\mathcal{A}$, $\Phi$ at $q\neq 0$. Furthermore, we introduce a generalized Katanin approximation to account for higher order couplings in the flow equations. We refer to Appendix~\ref{app: flow eqs} for a derivation of the latter. We now show that our reduced set of flow equations for the various couplings can be integrated. We first focus on the longitudinal channel, while in the transverse one the flow equations possess the same structure. 

\begin{figure}[t]
    \centering
    \includegraphics[width=0.32\textwidth]{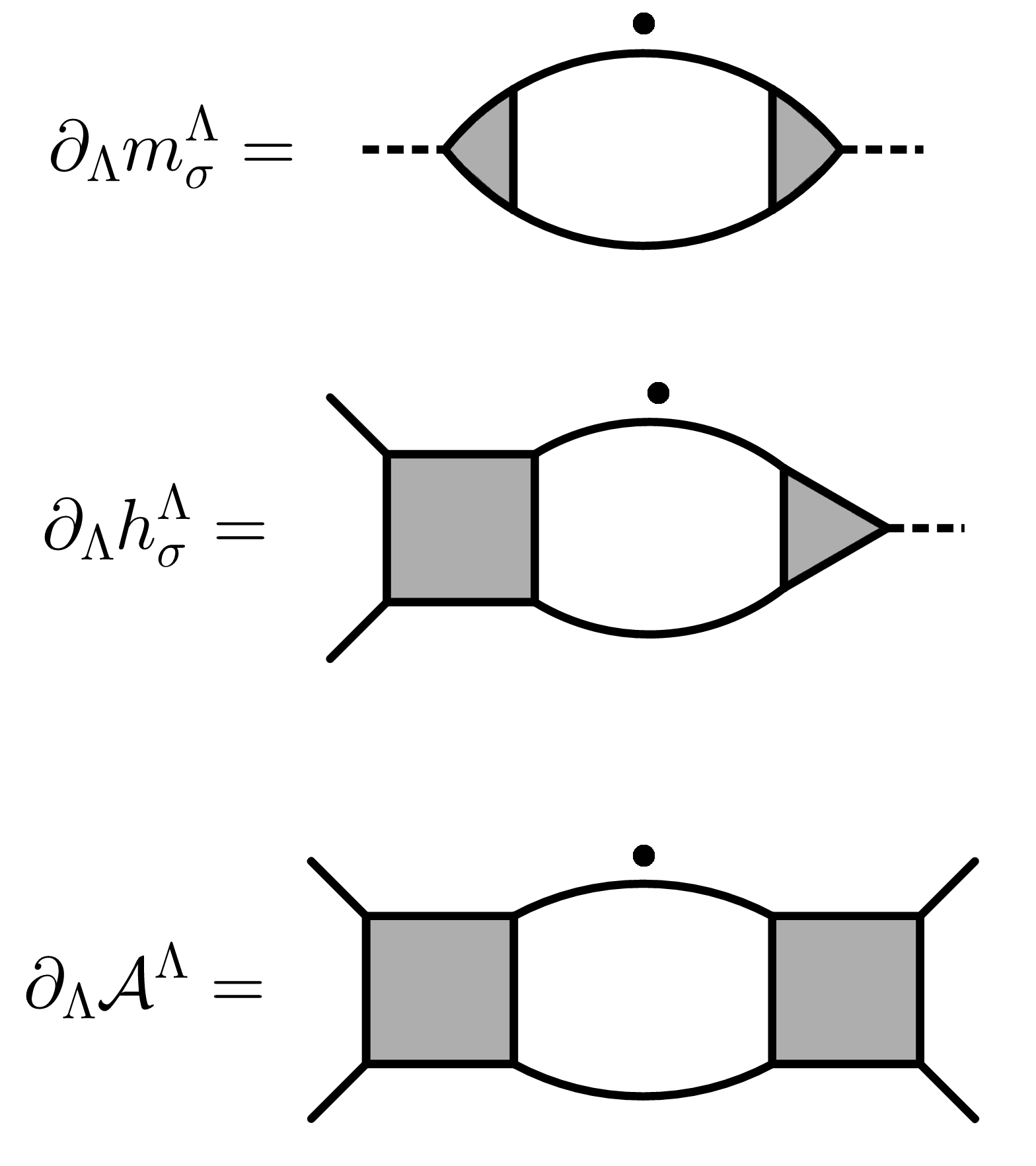}
    \caption{Schematic representation of flow equations for the mass and the couplings in the longitudinal channel. Full lines represent Nambu matrix propagators, triangles the Yukawa coupling $h_\sigma$ and squares the residual interaction $\mathcal{A}$. The black dots over fermionic legs represent full derivatives with respect to the scale $\Lambda$.}
    \label{fig: flow eqs}
\end{figure}

The flow equation for the longitudinal bosonic mass (inverse propagator at $q=0$) reads
\begin{equation}
    \begin{split}
        \partial_\Lambda m_\sigma^\Lambda=&\int_{k,k'} h^\Lambda_\sigma(k) \left[\partial_\Lambda\Pi^\Lambda_{11}(k,k')\right] h^\Lambda_\sigma(k') \\
        \equiv &\left[h^\Lambda_\sigma\right]^T\left[\partial_\Lambda\Pi^\Lambda_{11}\right] h^\Lambda_\sigma.
    \end{split}
    \label{eq: flow P sigma}
\end{equation}
Similarly, the equation for the longitudinal Yukawa coupling is
\begin{equation}
    \partial_\Lambda h^\Lambda_\sigma=\mathcal{A}^\Lambda\left[\partial_\Lambda\Pi^\Lambda_{11}\right]h^\Lambda_\sigma,
    \label{eq: flow h sigma}
\end{equation}
and the one for the residual two fermion longitudinal interaction is given by
\begin{equation}
    \partial_\Lambda\mathcal{A}^\Lambda=\mathcal{A}^\Lambda\left[\partial_\Lambda\Pi^\Lambda_{11}\right]\mathcal{A}^\Lambda.
    \label{eq: A flow eq}
\end{equation}
The above flow equations are pictorially shown in Fig.~\ref{fig: flow eqs}. The initial conditions at $\Lambda=\Lambda_s$ read, for both channels,
\begin{equation}
    \begin{split}
        &m_\sigma^{\Lambda_s}=m_\pi^{\Lambda_s}=m^{\Lambda_s}\\
        &h_\sigma^{\Lambda_s}=h_\pi^{\Lambda_s}=h^{\Lambda_s}\\
        &\mathcal{A}^{\Lambda_s}=\Phi^{\Lambda_s}=\mathcal{Q}^{\Lambda_s}.
    \end{split}
\end{equation}
We start by integrating the equation for the residual two fermion longitudinal interaction $\mathcal{A}$. Eq.~\eqref{eq: A flow eq} can be solved exactly as we have done in the fermionic formalism, obtaining for $\mathcal{A}$
\begin{equation}
    \mathcal{A}^\Lambda = \left[1-\widetilde{\mathcal{Q}}^{\Lambda_s}\Pi_{11}^\Lambda\right]^{-1}\widetilde{\mathcal{Q}}^{\Lambda_s},
    \label{eq: A}
\end{equation}
where we have introduced a reduced residual two fermion interaction $\widetilde{\mathcal{Q}}$
\begin{equation}
    \widetilde{\mathcal{Q}}^{\Lambda_s}=\left[1+\mathcal{Q}^{\Lambda_s}\Pi^{\Lambda_s}\right]^{-1}\mathcal{Q}^{\Lambda_s}.
    \label{eq: reduced C tilde}
\end{equation}
We are now in the position to employ this result and plug it in Eq.~\eqref{eq: flow h sigma} for the Yukawa coupling. The latter can be integrated as well. Its solution reads 
\begin{equation}
    h_\sigma^\Lambda= \left[1-\widetilde{\mathcal{Q}}^{\Lambda_s}\Pi_{11}^\Lambda\right]^{-1}\widetilde{h}^{\Lambda_s},
    \label{eq: h_sigma}
\end{equation}
where the introduction of a "reduced" Yukawa coupling
\begin{equation}
    \widetilde{h}^{\Lambda_s}=\left[1+\mathcal{Q}^{\Lambda_s}\Pi^{\Lambda_s}\right]^{-1}h^{\Lambda_s}
    \label{eq: reduced yukawa}
\end{equation}
is necessary. This Bethe-Salpeter-like equation for the Yukawa coupling is similar in structure to the parquetlike equations for the three-leg vertex derived in Ref.~\cite{Krien2019_II}.
Finally, we can use the two results of Eqs.~\eqref{eq: A} and~\eqref{eq: h_sigma} and plug them in the equation for the bosonic mass, whose integration provides
\begin{equation}
    m_\sigma^\Lambda=\widetilde{m}^{\Lambda_s}-\left[\widetilde{h}^{\Lambda_s}\right]^T\Pi_{11}^\Lambda\,h_\sigma^\Lambda,
    \label{eq: P_sigma}
\end{equation}
where, by following definitions introduced above, the "reduced" bosonic mass is given by
\begin{equation}
    \widetilde{m}^{\Lambda_s}=m^{\Lambda_s}+\left[\widetilde{h}^{\Lambda_s}\right]^T\Pi^{\Lambda_s}\,h^{\Lambda_s}.
    \label{eq: reduced mass P tilde}
\end{equation}
In the transverse channel, the equations have the same structure and can be integrated in the same way. Their solutions read
\begin{align}
    &\Phi^\Lambda = \left[1-\widetilde{\mathcal{Q}}^{\Lambda_s}\Pi_{22}^\Lambda\right]^{-1}\widetilde{\mathcal{Q}}^{\Lambda_s},
    \label{eq: Phi}\\
    &h_\pi^\Lambda= \left[1-\widetilde{\mathcal{Q}}^{\Lambda_s}\Pi_{22}^\Lambda\right]^{-1}\widetilde{h}^{\Lambda_s},
    \label{eq: h_pi}\\
    &m_\pi^\Lambda=\widetilde{m}^{\Lambda_s}-\left[\widetilde{h}^{\Lambda_s}\right]^T\Pi_{22}^\Lambda\,h_\pi^\Lambda.
    \label{eq: goldstone mass}
\end{align}
Eq.~\eqref{eq: goldstone mass} provides the mass of the transverse mode, which, according to the Goldstone theorem, must be zero. We will show later that this is indeed fulfilled. 

It is worthwhile to point out that the combinations
\begin{equation}
    \begin{split}
        &\frac{h_\sigma^\Lambda \left[h_\sigma^\Lambda\right]^T}{m_\sigma^\Lambda}+\mathcal{A}^\Lambda\\
        &\frac{h_\pi^\Lambda \left[h_\pi^\Lambda\right]^T}{m_\pi^\Lambda}+\Phi^\Lambda
    \end{split}
    \label{eq: eff fer interactions}
\end{equation}
obey the same flow equations, Eqs.~\eqref{eq: flow eq Va fermionic} and~\eqref{eq: flow eq Vphi fermionic}, as the vertices in the fermionic formalism and share the same initial conditions. Therefore the solutions for these quantities coincide with expressions~\eqref{eq: Va solution fermionic} and~\eqref{eq: Vphi solution fermionic}, respectively. Within this equivalence, it is interesting to express the irreducible vertex $\widetilde{V}^{\Lambda_s}$ of Eq.~\eqref{eq: irr vertex fermionic} in terms of the quantities, $\mathcal{Q}^{\Lambda_s}$, $h^{\Lambda_s}$ and $m^{\Lambda_s}$, introduced in the factorization in Eq.~\eqref{eq: vertex at Lambda crit}:
\begin{equation}
    \widetilde{V}^{\Lambda_s}=\frac{\widetilde{h}^{\Lambda_s}\left[\widetilde{h}^{\Lambda_s}\right]^T}{\widetilde{m}^{\Lambda_s}}+\widetilde{\mathcal{Q}}^{\Lambda_s},
    \label{eq: irr V bosonic formalism}
\end{equation}
where $\widetilde{\mathcal{Q}}^{\Lambda_s}$, $\widetilde{h}^{\Lambda_s}$ and $\widetilde{m}^{\Lambda_s}$ were defined in Eqs.~\eqref{eq: reduced C tilde},~\eqref{eq: reduced yukawa} and~\eqref{eq: reduced mass P tilde}. For a proof see Appendix~\ref{app: irr V bosonic formalism}. Relation~\eqref{eq: irr V bosonic formalism} is of particular interest because it states that when the full vertex is expressed as in Eq.~\eqref{eq: vertex at Lambda crit}, then the irreducible one will obey a similar decomposition, where the bosonic propagator, Yukawa coupling and residual two fermion interaction are replaced by their "reduced" counterparts. This relation holds even for $q\neq 0$.
\subsection{Ward identity for the gap and Goldstone theorem}
\begin{figure}[t]
    \centering
    \includegraphics[width=0.35\textwidth]{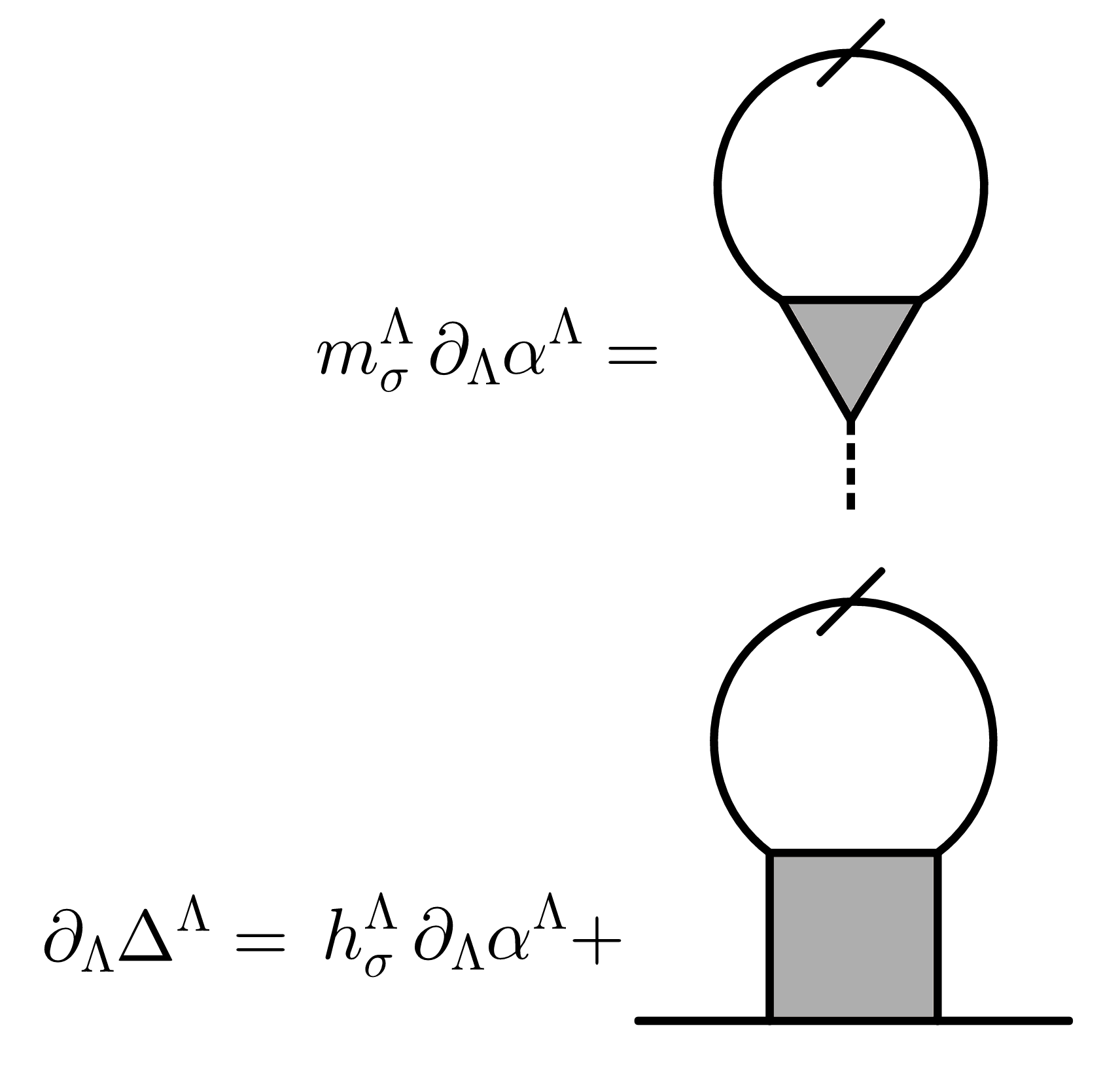}
    \caption{Schematic representation of flow equations for the bosonic expectation value $\alpha^\Lambda$ and fermionic gap $\Delta^\Lambda$. Besides the slashed lines, representing Nambu matrix propagators with a scale derivative acting only on the regulator, the conventions for the symbols are the same as in Fig.~\ref{fig: flow eqs}.}
    \label{fig: flow eqs gaps}
\end{figure}

We now focus on the flow of the fermionic gap and the bosonic expectation value and express a relation that connects them. Their flow equations are given by (see Appendix~\ref{app: flow eqs}) 
\begin{equation}
    \partial_\Lambda \alpha^\Lambda=\frac{1}{m_\sigma^\Lambda}\left[h_\sigma^\Lambda\right]^T\widetilde{\partial}_\Lambda F^\Lambda,
    \label{eq: dalpha dLambda main text}
\end{equation}
and
\begin{equation}
    \begin{split}
        \partial_\Lambda \Delta^\Lambda &= \partial_\Lambda \alpha^\Lambda\, h_\sigma^\Lambda+\mathcal{A}^\Lambda\widetilde{\partial}_\Lambda F^\Lambda\\
        &= \left[\frac{h_\sigma^\Lambda \left[h_\sigma^\Lambda\right]^T}{m_\sigma^\Lambda}+\mathcal{A}^\Lambda\right]\widetilde{\partial}_\Lambda F^\Lambda,
        \label{eq: gap eq main text}
    \end{split}
\end{equation}
with $F^\Lambda$ given by Eq.~\eqref{eq: F definition}. In Fig.~\ref{fig: flow eqs gaps} we show a pictorial representation.
Eq.~\eqref{eq: dalpha dLambda main text} can be integrated, with the help of the previously obtained results for $\mathcal{A}$, $h_\sigma$ and $m_\sigma$, yielding
 \begin{equation}
    \alpha^\Lambda=\frac{1}{\widetilde{m}^{\Lambda_s}}\left[\widetilde{h}^{\Lambda_s}\right]^T F^\Lambda.
    \label{eq: alpha solution}
\end{equation}
In the last line of Eq.~\eqref{eq: gap eq main text}, as previously discussed, the object in angular brackets equals the full vertex $V_\mathcal{A}$ of the fermionic formalism. Thus, integration of the gap equation is possible and the result is simply Eq.~\eqref{eq: gap equation fermionic} of the fermionic formalism. However, if we now insert the expression in Eq.~\eqref{eq: irr V bosonic formalism} for the irreducible vertex within the "fermionic" form (Eq.~\eqref{eq: gap equation fermionic}) of the gap equation, and use relation~\eqref{eq: Pi22=F/delta}, we get:
\begin{equation}
    \Delta^\Lambda(k)=\alpha^\Lambda h_\pi^\Lambda(k).
    \label{eq: Ward Identity}
\end{equation}
This equation is the Ward identity for the mixed boson-fermion system related to the global U(1) symmetry~\cite{Obert2013}. In Appendix~\ref{app: loop} we propose a self consistent loop for the calculation of $\alpha$, $h_{\pi}$, through Eqs.~\ref{eq: alpha solution} and~\ref{eq: h_pi}, and subsequently the superfluid gap $\Delta$.
Let us now come back to the question of the Goldstone theorem. For the mass of the Goldstone boson to be zero, it is necessary for Eq.~\eqref{eq: goldstone mass} to vanish. We show that this is indeed the case. With the help of Eq.~\eqref{eq: Pi22=F/delta}, we can reformulate the equation for the transverse mass in the form 
\begin{equation}
    \begin{split}
        m^\Lambda_\pi &= \widetilde{m}^{\Lambda_s}-\int_k \widetilde{h}^{\Lambda_s}(k)F^\Lambda(k)\frac{h^\Lambda_\pi(k)}{\Delta^\Lambda(k)}\\
        &=\widetilde{m}^{\Lambda_s}-\frac{1}{\alpha^{\Lambda}}\int_k \widetilde{h}^{\Lambda_s}(k)F^\Lambda(k),
    \end{split}
\end{equation}
where the Ward Identity $\Delta=\alpha h_\pi$ was applied in the last line. We see that the expression for the Goldstone boson mass vanishes when $\alpha$ obeys its self consistent equation, Eq.~\eqref{eq: alpha solution}. This proves that our truncation of flow equations fulfills the Goldstone theorem. \\
\hgl{Constructing a truncation of the fRG flow equations which fulfills the Ward identities and the Goldstone theorem is, in general, a nontrivial task. In Ref.~\cite{Bartosch2009}, in which the order parameter fluctuations have been included on top of the Hartree-Fock solution, no distinction has been made between the longitudinal and transverse Yukawa couplings and the Ward identity~\eqref{eq: Ward Identity} as well as the Goldstone theorem have been enforced by construction, by calculating the gap and the bosonic expectation values from them rather than from their flow equations. Similarly, in Ref.~\cite{Obert2013}, in order for the flow equations to fulfill the Goldstone theorem, it was necessary to impose $h_\sigma=h_\pi$ and use only the flow equation of $h_\pi$ for both Yukawa couplings. Within the present approach, due to the mean-field-like nature of the truncation, the Ward identity~\eqref{eq: Ward Identity} and the Goldstone theorem are automatically fulfilled by the flow equations.}
\subsection{Equivalence of bosonic and fermionic formalisms}
\label{subsec: equivalence bos and fer}
As we have proven in the previous sections, within the MF approximation the fully fermionic formalism of Sec.~\ref{sect: fermionic formalism} and the bosonized approach introduced in the present section provide the same results for the superfluid gap and for the effective two fermion interactions. 
Notwithstanding the formal equivalence, the bosonic formulation relies on a further requirement. In Eqs.~\eqref{eq: Phi} and~\eqref{eq: h_pi} we assumed the matrix $\left[1-\widetilde{\mathcal{Q}}^{\Lambda_s}\Pi_{22}^\Lambda\right]$ to be invertible. This statement is exactly equivalent to assert that the two fermion residual interaction $\Phi$ remains finite. Otherwise the Goldstone mode would lie in this coupling and not (only) in the Hubbard-Stratonovich boson. This fact cannot occur if the flow is stopped at a scale $\Lambda_s$ coinciding with the critical scale $\Lambda_c$ at which the (normal) bosonic mass $m^\Lambda$ turns zero, but it could take place if one considers symmetry breaking in more than one channel. In particular, if one allows the system to develop two different orders and stops the flow when the mass of one of the two associated bosons becomes zero, it could happen that, within a MF approximation for both order types, the appearance of a finite gap in the first channel makes the two fermion transverse residual interaction in the other channel diverging. In that case one can apply the technique of the \textit{flowing bosonization}~\cite{Friederich2010,Friederich2011}, by reassigning to the bosonic sector the (most singular part of the) two fermion interactions that are generated during the flow. It can be proven that also this approach gives the same results for the gap and the effective fermionic interactions in Eq.~\eqref{eq: eff fer interactions} as the fully fermionic formalism. 
\section{Vertex bosonization}
\label{sec: vertex bosonization}
In this section we present a systematic procedure to extract the quantities in Eq.~\eqref{eq: vertex at Lambda crit} from a given vertex, within an approximate framework. The full vertex in the symmetric phase can be written as~\cite{Husemann2009,Husemann2012}
\begin{equation}
    \begin{split}
        V^\Lambda(k_1,k_2,k_3)&=V^{\Lambda_\text{ini}}(k_1,k_2,k_3)\\
        &+\phi^\Lambda_p\left(k_1,k_3;k_1+k_2\right)\\
        &-\phi^\Lambda_m\left(k_1,k_2;k_2-k_3\right)\\
        &-\frac{1}{2}\phi^\Lambda_m\left(k_1,k_2;k_3-k_1\right)\\
        &+\frac{1}{2}\phi^\Lambda_c\left(k_1,k_2;k_3-k_1\right),
    \end{split}
    \label{eq: channel decomposition}
\end{equation}
where $V^{\Lambda_\text{ini}}$ is the vertex at the initial scale, and we call $\phi_p$ pairing channel, $\phi_m$ magnetic channel and $\phi_c$ charge channel. Each of this functions depends on a bosonic and two fermionic variables.
Within the so called 1-loop approximation, where one neglects the 3-fermion coupling in Eq.~\eqref{eq: vertex flow equation symm}, in the Katanin scheme~\cite{Katanin2004}, or in more involved schemes, such as the 2-loop~\cite{Eberlein2014} or the multiloop~\cite{Kugler2018_I,Kugler2018_II}, one is able to assign one or more of the terms of the flow equation~\eqref{eq: vertex flow equation symm} for $V^\Lambda$ to each of the channels, in a way that their last bosonic argument enters only parametrically in the formulas. This is the reason why the decomposition in Eq.~\eqref{eq: channel decomposition} is useful. The vertex at the initial scale can be set equal to the bare (sign-reversed) Hubbard interaction $-U$ in a weak-coupling approximation, or as in the recently introduced DMF\textsuperscript{2}RG scheme, to the vertex computed via DMFT~\cite{Taranto2014,Vilardi2019}.

In order to simplify the treatment of the dependence on fermionic spatial momenta of the various channels, one often introduces a complete basis of Brillouin zone form factors $\{f^\ell_\mathbf{k}\}$ and expands each channel in this basis~\cite{Lichtenstein2017}
\begin{equation}
    \begin{split}
        \phi^\Lambda_X(k,k';q)=\sum_{\ell\ell'} \phi^{\Lambda}_{X,\ell\ell'}(\nu,\nu';q)f^\ell_{\mathbf{k}+(\text{sgn}X)\mathbf{q}/2}\,f^{\ell'}_{\mathbf{k'}-\mathbf{q}/2},
    \end{split}
    \label{eq: form factor expansion}
\end{equation}
with $X=p$, $m$ or $c$, and $\text{sgn}\,p=-1$, $\text{sgn}\,c=\text{sgn}\,m=+1$. For practical calculations the above sum is truncated to a finite number of form factors and often only diagonal terms, $\ell=\ell'$, are considered. Within the form factor truncated expansion, one is left with the calculation of a finite number of channels that depend on a bosonic "$d$+1 momentum" $q=(\mathbf{q},\Omega)$ and two fermionic Matsubara frequencies $\nu$ and $\nu'$.

We will now show how to obtain the decomposition introduced in Eq.~\eqref{eq: vertex at Lambda crit} within the form factor expansion.
We focus on only one of the channels in Eq.~\eqref{eq: channel decomposition}, depending on the type of order we are interested in, and factorize its dependence on the two fermionic Matsubara frequencies. We introduce the so called channel asymptotics, that is the functions that describe the channels for large $\nu$, $\nu'$. From now on we adopt the shorthand $\lim_{\nu\rightarrow\infty}g(\nu)=g(\infty)$ for whatever $g$, function of $\nu$. By considering only diagonal terms in the form factor expansion in Eq.~\eqref{eq: form factor expansion}, we can write the channels as~\cite{Wentzell2016}:
\begin{equation}
    \begin{split}
        \phi_{X,\ell}^\Lambda(\nu,\nu';q)&=\mathcal{K}_{X,\ell}^{(1)\Lambda}(q)+\mathcal{K}_{X,\ell}^{(2)\Lambda}(\nu;q)\\
        &+\overline{\mathcal{K}}_{X,\ell}^{(2)\Lambda}(\nu';q)
        +\delta\phi^\Lambda_{X,\ell}(\nu,\nu';q),    
    \end{split}
    \label{eq: vertex asymptotics}
\end{equation}
with
\begin{equation}
    \begin{split}
        &\mathcal{K}_{X,\ell}^{(1)\Lambda}(q)=\phi_{X,\ell}^\Lambda(\infty,\infty;q)\\
        &\mathcal{K}_{X,\ell}^{(2)\Lambda}(\nu;q)=\phi_{X,\ell}^\Lambda(\nu,\infty;q)-\mathcal{K}_{X,\ell}^{(1)\Lambda}(q)\\
        &\overline{\mathcal{K}}_{X,\ell}^{(2)\Lambda}(\nu';q)=\phi_{X,\ell}^\Lambda(\infty,\nu';q)-\mathcal{K}_{X,\ell}^{(1)\Lambda}(q)\\
        &\delta\phi^\Lambda_{X,\ell}(\nu,\infty;q)=\delta\phi^\Lambda_{X,\ell}(\infty,\nu';q)=0.
    \end{split}
    \label{eq: asymptotics properties}
\end{equation}
According to Ref.~\cite{Wentzell2016}, these functions are related to physical quantities. $\mathcal{K}_{X,\ell}^{(1)}$ turns out to be proportional to the relative susceptibility and the combination $\mathcal{K}_{X,\ell}^{(1)}+\mathcal{K}_{X,\ell}^{(2)}$ (or $\mathcal{K}_{X,\ell}^{(1)}+\overline{\mathcal{K}}_{X,\ell}^{(2)}$) to the so called boson-fermion vertex, that describes both the response of the Green's function to an external field~\cite{VanLoon2018} and the coupling between a fermion and an effective boson. In principle one should be able to calculate the above quantities diagrammatically (see Ref.~\cite{Wentzell2016} for the details) without performing any limit. However, it is well known how fRG truncations, in particular the 1-loop approximation, do not properly weight all the Feynman diagrams contributing to the vertex, so that the diagrammatic calculation and the high frequency limit give two different results. To keep the property in the last line of Eq.~\eqref{eq: asymptotics properties}, we choose to perform the limits. We rewrite Eq.~\eqref{eq: vertex asymptotics} in the following way:
\begin{equation}
    \begin{split}
        &\phi_{X,\ell}^\Lambda(\nu,\nu';q)=\\
        &=\frac{\left[\mathcal{K}_{X,\ell}^{(1)\Lambda}+\mathcal{K}_{X,\ell}^{(2)\Lambda}\right]\left[\mathcal{K}_{X,\ell}^{(1)\Lambda}+\overline{\mathcal{K}}_{X,\ell}^{(2)\Lambda}\right]}{\mathcal{K}_{X,\ell}^{(1)\Lambda}}+\mathcal{R}_{X,\ell}^\Lambda\\
        &=\frac{\phi_{X,\ell}^\Lambda(\nu,\infty;q)\phi_{X,\ell}^\Lambda(\infty,\nu';q)}{\phi_{X,\ell}^\Lambda(\infty,\infty;q)}+\mathcal{R}_{X,\ell}^\Lambda(\nu,\nu';q),
    \end{split}
    \label{eq: vertex separation}
\end{equation}
where we have made the frequency and momentum dependencies explicit only in the second line and we have defined
\begin{equation}
    \mathcal{R}_{X,\ell}^\Lambda(\nu,\nu';q)=\delta\phi^\Lambda_{X,\ell}(\nu,\nu';q)-\frac{\mathcal{K}_{X,\ell}^{(2)\Lambda}(\nu;q)\overline{\mathcal{K}}_{X,\ell}^{(2)\Lambda}(\nu';q)}{\mathcal{K}_{X,\ell}^{(1)\Lambda}(q)}.
\end{equation}
From the definitions given above, it is obvious that the rest function $\mathcal{R}_{X,\ell}$ decays to zero in all frequency directions.

Since the first term of Eq.~\eqref{eq: vertex separation} is separable by construction, we choose to identify this term with the first one of Eq.~\eqref{eq: vertex at Lambda crit}. Indeed, in many cases the vertex divergence is manifest already in the asymptotic $\mathcal{K}_{X,\ell}^{(1)\Lambda}$, that we recall to be proportional to the susceptibility of the channel. There are however situations in which the functions $\mathcal{K}^{(1)}$ and $\mathcal{K}^{(2)}$ are zero even close to an instability in the channel, an important example being the d-wave superconducting instability in the repulsive Hubbard model. In general, this occurs for those channels that, within a Feynman diagram expansion, cannot be constructed with a ladder resummation with the bare vertex. In the Hubbard model, due to the locality of the bare interaction, this happens for every $\ell\neq 0$, that is for every term in the form factor expansion different than the s-wave contribution. In this case one should adopt a different approach and, for example, replace the limits to infinity in Eq.~\eqref{eq: vertex separation} by some given values of the Matsubara frequencies, $\pm \pi T$ for example.
\section{Results for the attractive Hubbard model at half-filling}
\label{sec: results}
In this section we report some exemplary results of the equations derived within the bosonic formalism, for the attractive two-dimensional Hubbard model. We focus on the half-filled case. For pure nearest neighbors hopping with amplitude $-t$, the band dispersion $\xi_\mathbf{k}$ is given by
\begin{equation}
    \xi_\mathbf{k} = - 2 t \left( \cos k_x + \cos k_y \right) -\mu,
    \label{eq: dispersion band}
\end{equation}
with $\mu=0$ at half-filling. We choose the onsite attraction and the temperature to be $U=-4t$ and $T=0.1t$. All results are presented in units of the hopping parameter $t$. 
\subsection{Symmetric phase}
In the symmetric phase, in order to run a fRG flow, we introduce the $\Omega$-regulator~\cite{Husemann2009}
\begin{equation}
    R^\Lambda(k) = \left(i\nu-\xi_\mathbf{k}\right) \frac{\Lambda^2}{\nu^2},
\end{equation}
so that the initial scale is $\Lambda_\text{init}=+\infty$ (fixed to a large number in the numerical calculation) and the final one $\Lambda_\text{fin}=0$. We choose a 1-loop truncation, that is we neglect the last term of Eq.~\eqref{eq: vertex flow equation symm}, and use the decomposition in Eq.~\eqref{eq: channel decomposition} with a form factor expansion. We truncate Eq.~\eqref{eq: form factor expansion} only to the first term, that is we use only s-wave, $f^{(0)}_\mathbf{k}\equiv 1$, form factors. Within these approximations, the vertex reads 
\begin{equation}
    \begin{split}
        V^\Lambda(&k_1,k_2,k_3) = - U + \mathcal{P}^{\Lambda}_{\nu_1\nu_3}(k_1+k_2) \\
        - &\mathcal{M}^{\Lambda}_{\nu_1\nu_2}(k_2-k_3)\\
        -&\frac{1}{2} \mathcal{M}^{\Lambda}_{\nu_1\nu_2}(k_3-k_1)
        +\frac{1}{2} \mathcal{C}^{\Lambda}_{\nu_1\nu_2}(k_3-k_1),
    \end{split}
    \label{eq: channel decomposition attractive model}
\end{equation}
where $\mathcal{P}$, $\mathcal{M}$, $\mathcal{C}$, are referred as pairing, magnetic and charge channel, respectively.
Furthermore, we focus only on the spin-singlet component of the pairing (the triplet one is very small in the present parameter region), so that we require the pairing channel to obey~\cite{Rohringer2012}
\begin{equation}
    \mathcal{P}^{\Lambda}_{\nu\nu'}(q) = \mathcal{P}^{\Lambda}_{\Omega-\nu,\nu'}(q) = \mathcal{P}^{\Lambda}_{\nu,\Omega-\nu'}(q),
\end{equation}
with $q=(\mathbf{q},\Omega)$.
The initial condition for the vertex reads
\begin{equation}
    V^{\Lambda_\text{init}}(k_1,k_2,k_3) = - U,
\end{equation}
so that $\mathcal{P}^{\Lambda_\text{init}}=\mathcal{M}^{\Lambda_\text{init}}=\mathcal{C}^{\Lambda_\text{init}}=0$.
Neglecting the fermionic self-energy, $\Sigma^\Lambda(k)\equiv0$, we run a flow for these three quantities until one (ore more) of them diverges. Under a technical point of view, each channel is computed by keeping 50 positive and 50 negative values for each of the three Matsubara frequencies (two fermionic, one bosonic) on which it depends. Frequency asymptotics are also taken into account, by following Ref.~\cite{Wentzell2016}. The momentum dependence of the channel is treated by discretizing with 38 patches the region $\mathcal{B}=\{(k_x,k_y): 0\leq k_y\leq k_x\leq\pi\}$ in the Brillouin zone and extending to the other regions by using lattice symmetries. The expressions of the flow equations are reported in Appendix~\ref{app: flow eqs symm phase}. 

\begin{figure}[t]
    \centering
    \includegraphics[width=0.45\textwidth]{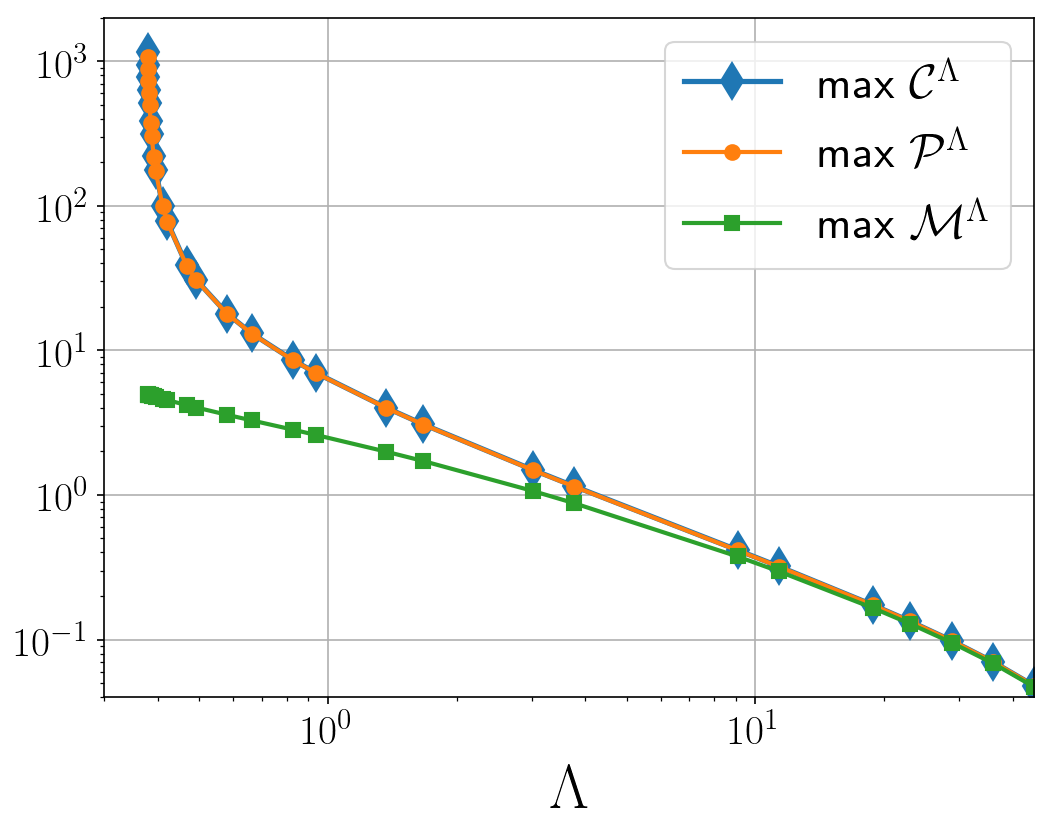}
    \caption{Flow of the maximum values of the pairing, magnetic and charge channel. The maximum value of the charge channel at zero frequency and momentum $(\pi,\pi)$ and the one for the pairing channel at $q=0$ coincide, within the numerical accuracy, and both exceed the threshold $10^3t$ at the stopping scale, signaling an instability to the formation of an order parameter given by any linear combination of the superfluid and the charge density wave ones.}
    \label{fig: flow channels}
\end{figure}

Due to particle-hole symmetry occurring at half-filling, pairing fluctuations at $\mathbf{q}=0$ combine with charge fluctuations at $\mathbf{q}=(\pi,\pi)$ to form an order parameter with SO(3) symmetry~\cite{Micnas1990}. Indeed, with the help of a canonical particle-hole transformation, one can map the attractive half-filled Hubbard model onto the repulsive one. Within this duality, the SO(3)-symmetric magnetic order parameter is mapped onto the above mentioned combined charge-pairing order parameter and vice versa. This is the reason why we find a critical scale, $\Lambda_c$, at which both $\mathcal{C}((\pi,\pi),0)$ and $\mathcal{P}(\mathbf{0},0)$ diverge, as shown in Fig.~\ref{fig: flow channels}. On a practical level, we define the stopping scale $\Lambda_s$ as the scale at which one (or more, in this case) channel exceeds $10^3t$. With our choice of parameters, we find that at $\Lambda_s \simeq 0.378t$ both $\mathcal{C}$ and $\mathcal{P}$ cross our threshold.
In the SSB phase, we choose to restrict the ordering to the pairing channel, thus excluding the formation of charge density waves. This choice is always possible because we have the freedom to choose the "direction" in which our order parameter points. Indeed, in the particle-hole dual repulsive model, our choice would be equivalent to choose the (antiferro-) magnetic order parameter to lie in the $xy$ plane. This choice is implemented by selecting the particle-particle channel as the only one contributing to the flow in the SSB phase, as exposed in Secs.~\ref{sect: fermionic formalism} and~\ref{sect: bosonic formalism}.

\begin{figure}[t]
    \centering
    \includegraphics[width=0.49\textwidth]{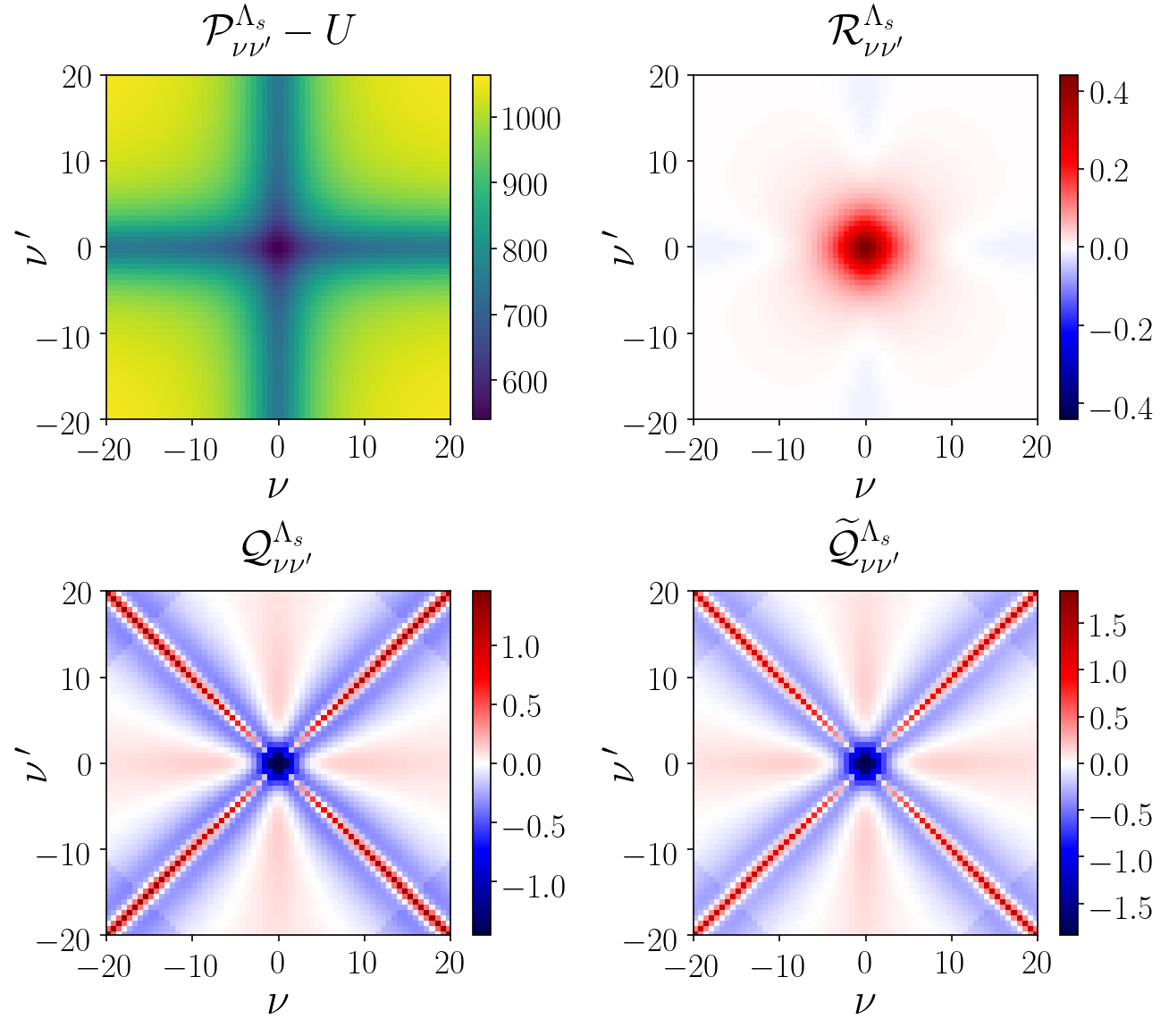}
    \caption{Couplings contributing to the total vertex at the stopping scale. \\
    \textit{Upper left}: pairing channel minus the bare interaction. At the stopping scale this quantity acquires very large values due to the vicinity to the pairing instability. \\
    \textit{Upper right}: rest function of the pairing channel minus the bare interaction. In the present regime the pairing channel is very well factorizable, giving rise to a small rest function.\\ 
    \textit{Lower left}: residual two fermion interaction. The choice of factorizing $\mathcal{P}^{\Lambda_s}-U$ instead of $\mathcal{P}^{\Lambda_s}$ alone makes the background of this quantity zero.\\
    \textit{Lower right}: reduced residual two fermion interaction. As well as the full one, this coupling has a zero background value, making calculations of couplings in the SSB phase more precise by reducing finite size effects in the matrix inversions.}
    
    \label{fig: vertices Lambda s}
\end{figure}

In order to access the SSB phase with our bosonic formalism, we need to perform the decomposition in Eq.~\eqref{eq: vertex at Lambda crit} for our vertex at $\Lambda_s$. Before proceeding, in order to be consistent with our form factor expansion in the SSB phase, we need to project $V$ in Eq.~\eqref{eq: channel decomposition attractive model} onto the s-wave form factors, because we want the quantities in the ordered phase to be functions of Matsubara frequencies only. Therefore we define the total vertex projected onto s-wave form factors 
\begin{equation}
    \overline{V}^{\Lambda_s}_{\nu\nu'}(q)=\int_{\mathbf{k},\mathbf{k}'}V^{\Lambda_s}\hskip -1mm\left(k,q-k,k'\right).
\end{equation}
Furthermore, since we are interested only in spin singlet pairing, we symmetrize it with respect to one of the two fermionic frequencies, so that in the end we are dealing with
\begin{equation}
    V^{\Lambda_s}_{\nu\nu'}(q)=\frac{\overline{V}^{\Lambda_s}_{\nu\nu'}(q)+\overline{V}^{\Lambda_s}_{\nu,\Omega-\nu'}(q)}{2}.
\end{equation}
In order to extract the Yukawa coupling $h^{\Lambda_s}$ and bosonic propagator $m^{\Lambda_s}$, we employ the strategy described in Sec.~\ref{sec: vertex bosonization}. Here, however, instead of factorizing the pairing channel $\mathcal{P}^{\Lambda_s}$ alone, we subtract from it the bare interaction $U$. In principle, $U$ can be assigned both to the pairing channel, to be factorized, or to the residual two fermion interaction, giving rise to the same results in the SSB phase. However, when in a real calculation the vertices are calculated on a finite frequency box, it is more convenient to have the residual two fermion interaction $\mathcal{Q}^{\Lambda_s}$ as small as possible, in order to reduce finite size effects in the matrix inversions needed to extract the reduced couplings in Eqs.~\eqref{eq: reduced C tilde},~\eqref{eq: reduced yukawa} and~\eqref{eq: reduced mass P tilde}, and in the calculation of $h_\pi$, in Eq.~\eqref{eq: h_pi}.  
Furthermore, since it is always possible to rescale the bosonic propagators and Yukawa couplings by a constant such that the vertex constructed with them (Eq.~\eqref{eq: vertex separation}) is invariant, we impose the normalization condition $h^{\Lambda_s}(\nu\rightarrow\infty;q)=1$.
In formulas, we have
\begin{equation}
    m^{\Lambda_s}(q)=\frac{1}{\mathcal{K}_{p,\ell=0}^{(1)\Lambda_s}(q)-U}=\frac{1}{\mathcal{P}^{\Lambda_s}_{\infty,\infty}(q)-U},
\end{equation}
and 
\begin{equation}
    \begin{split}
        h^{\Lambda_s}(\nu;q)&=\frac{\mathcal{K}_{p,\ell=0}^{(2)\Lambda_s}(\nu;q)+\mathcal{K}_{p,\ell=0}^{(1)\Lambda_s}(q)-U}{\mathcal{K}_{p,\ell=0}^{(1)\Lambda_s}(q)-U}\\
        &=\frac{\mathcal{P}^{\Lambda_s}_{\nu,\infty}(q)-U}{\mathcal{P}^{\Lambda_s}_{\infty,\infty}(q)-U}.
    \end{split}
\end{equation}
The limits are numerically performed by evaluating the pairing channel at large values of the fermionic frequencies.
The extraction of the factorizable part from the pairing channel minus the bare interaction defines the rest function
\begin{equation}
    \mathcal{R}^{\Lambda_s}_{\nu\nu'}(q)=\mathcal{P}^{\Lambda_s}_{\nu\nu'}(q)-U-\frac{h^{\Lambda_s}(\nu;q)h^{\Lambda_s}(\nu';q)}{m^{\Lambda_s}(q)},
\end{equation}
and the residual two fermion interaction $\mathcal{Q}$
\begin{equation}
    \begin{split}
        \mathcal{Q}^{\Lambda_s}_{\nu\nu'}(q)=&\left[V^{\Lambda_s}_{\nu\nu'}(q)-\mathcal{P}^{\Lambda_s}_{\nu\nu'}(q)+U\right]+\mathcal{R}^{\Lambda_s}_{\nu\nu'}(q)\\
        =&V^{\Lambda_s}_{\nu\nu'}(q)-\frac{h^{\Lambda_s}(\nu;q)h^{\Lambda_s}(\nu';q)}{m^{\Lambda_s}(q)}.
    \end{split}
\end{equation}
We are now in the position to extract the reduced couplings, $\widetilde{\mathcal{Q}}^{\Lambda_s}$, $\widetilde{h}^{\Lambda_s}$ and $\widetilde{m}^{\Lambda_s}$, defined in Eqs.~\eqref{eq: reduced C tilde},~\eqref{eq: reduced yukawa},~\eqref{eq: reduced mass P tilde}. This is achieved by numerically inverting the matrix (we drop the $q$-dependence from now on, assuming always $q=0$)
\begin{equation}
    \delta_{\nu\nu'} + \mathcal{Q}^{\Lambda_s}_{\nu\nu'}\, \chi^{\Lambda_s}_{\nu'}, 
\end{equation}
with
\begin{equation}
    \chi^{\Lambda_s}_{\nu} = T\int_{\mathbf{k}}G_0^{\Lambda_s}(k)G_0^{\Lambda_s}(-k), 
\end{equation}
and
\begin{equation}
    G_0^{\Lambda_s}(k) = \frac{1}{i\nu-\xi_\mathbf{k}+R^{\Lambda_s}(k)} =\frac{\nu^2}{\nu^2+\Lambda_s^2}\frac{1}{i\nu-\xi_\mathbf{k}}.
\end{equation}
In Fig.~\ref{fig: vertices Lambda s} we show the results for the pairing channel minus the bare interaction, the rest function, the residual two fermion interaction $\mathcal{Q}$ and the reduced one $\widetilde{\mathcal{Q}}$ at the stopping scale. One can see that in the present parameter region the pairing channel (minus $U$) is highly factorizable. Indeed, despite the latter being very large because of the vicinity to the instability, the rest function $\mathcal{R}$ remains very small, a sign that the pairing channel is well described by the exchange of a single boson. Furthermore, thanks to our choice of assigning the bare interaction to the factorized part, as we see in Fig.~\ref{fig: vertices Lambda s}, both $\mathcal{Q}$ and $\widetilde{\mathcal{Q}}$ possess frequency structures that arise from a background that is zero. 
Finally, the full bosonic mass at the stopping scale is close to zero, $m^{\Lambda_s}\simeq10^{-3} $, due to the vicinity to the instability, while the reduced one is finite, $\widetilde{m}^{\Lambda_s}\simeq 0.237$.
\subsection{SSB Phase}
In the SSB phase, instead of running the fRG flow, we employ the analytical integration of the flow equations described in Sec.~\ref{sect: bosonic formalism}. On a practical level, we implement a solution of the loop described in Appendix~\ref{app: loop}, that allows for the calculation of the bosonic expectation value $\alpha$, the transverse Yukawa coupling $h_\pi$ and subsequently the fermionic gap $\Delta$ through the Ward identity $\Delta=\alpha h_\pi$. In this section we drop the dependence on the scale, since we have reached the final scale $\Lambda_\text{fin}=0$. Note that, as exposed previously, in the half-filled attractive Hubbard model the superfluid phase sets in by breaking a SO(3) rather than a U(1) symmetry. This means that one should expect the appearance of two massless Goldstone modes. Indeed, besides the Goldstone boson present in the (transverse) particle-particle channel, another one appears in the particle-hole channel and it is related to the divergence of the charge channel at momentum $(\pi,\pi)$. However, within our choice of considering only superfluid order and within the MF approximation, this mode is decoupled from our equations.

\begin{figure}[t]
    \centering
    \includegraphics[width=0.45\textwidth]{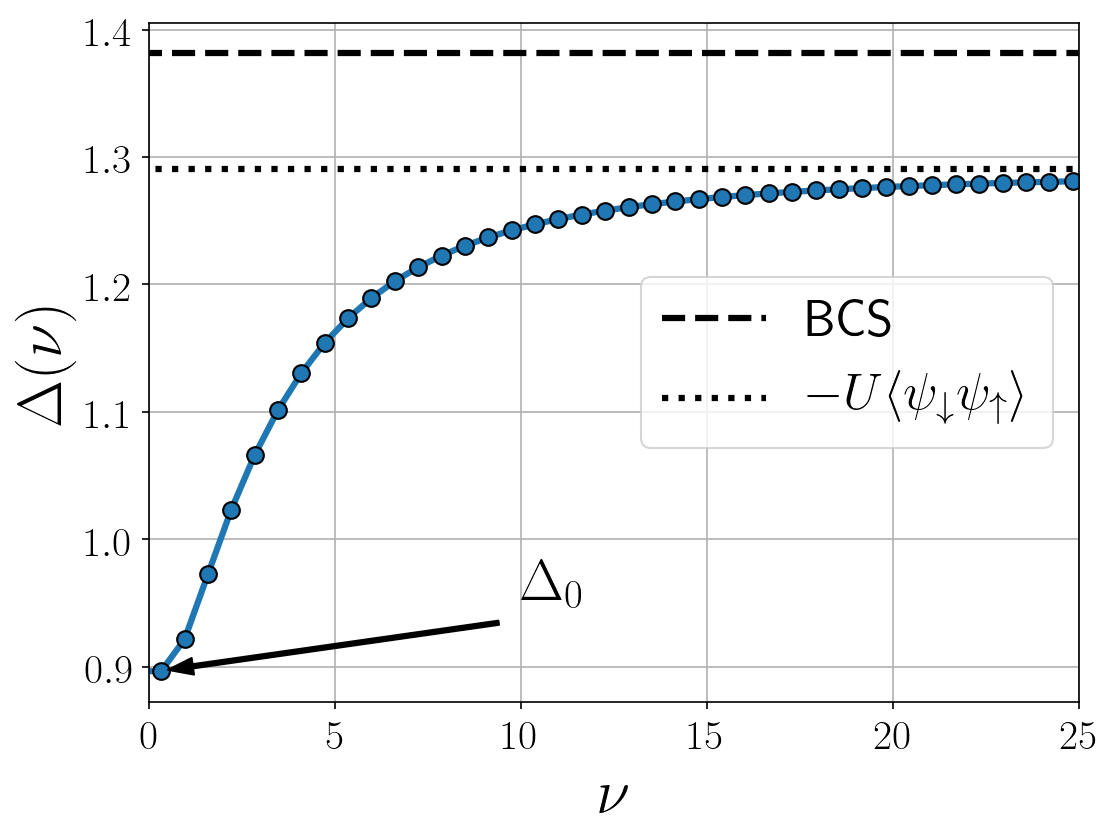}
    \caption{Frequency dependence of the superfluid gap. It interpolates between its value at the Fermi level, $\Delta_0$, and its asymptotic one. The dashed line marks the BCS value, while the dotted one  $-U$ times the Cooper pair expectation value.}
    \label{fig: gap}
\end{figure}

\begin{figure}[t]
    \centering
    \includegraphics[width=0.49\textwidth]{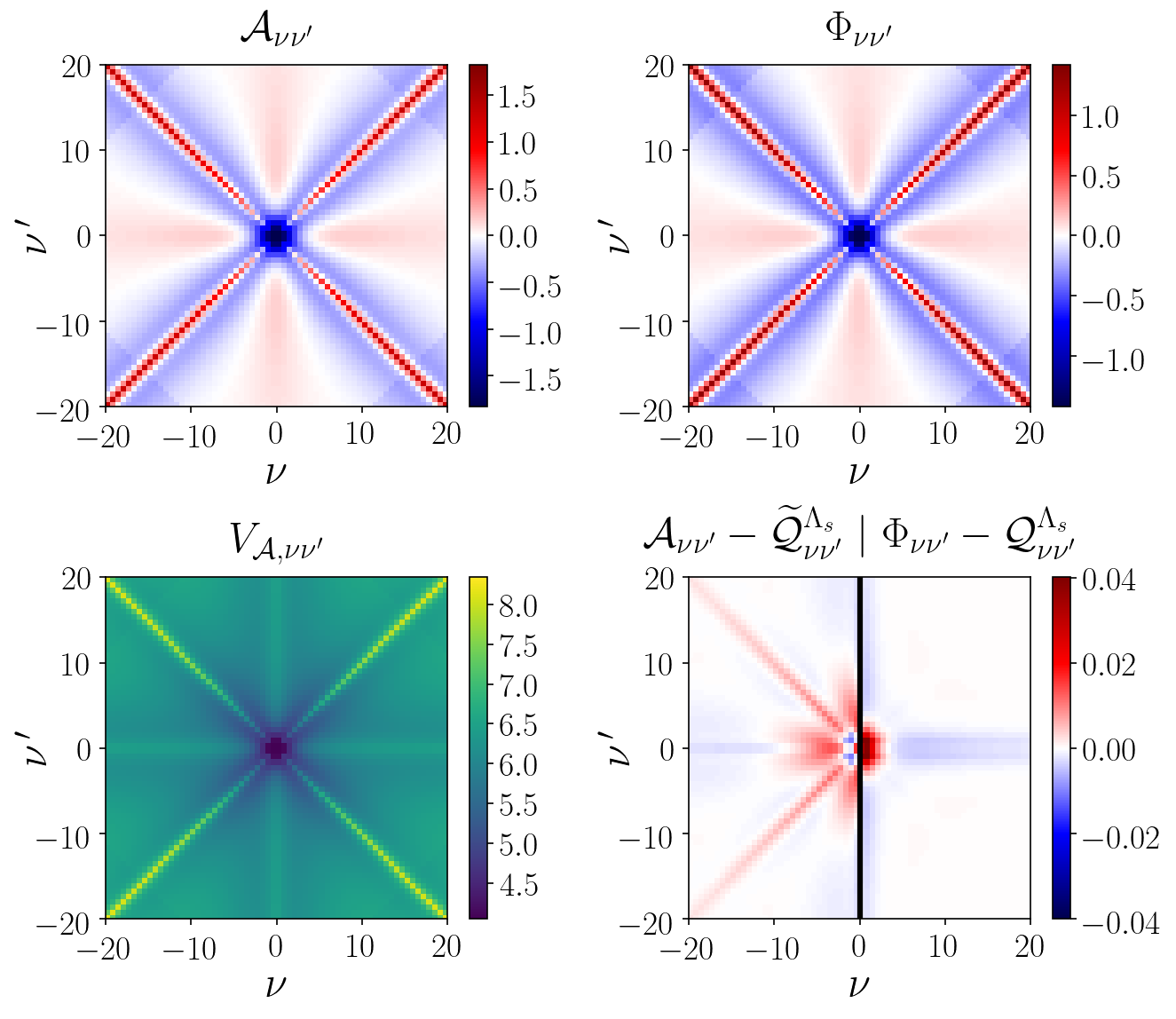}
    \caption{Effective interactions calculated in the SSB phase as functions of Matsubara frequencies.\\
    \textit{Upper left}: longitudinal residual two fermion interaction $\mathcal{A}$.\\
    \textit{Upper right}: transverse residual two fermion interaction $\Phi$.\\
    \textit{Lower left}: longitudinal effective two fermion interaction $V_\mathcal{A}$.\\
    \textit{Lower right}: longitudinal residual two fermion interaction $\mathcal{A}$ with its reduced counterpart $\widetilde{\mathcal{Q}}$ at the stopping scale subtracted (left), and transverse longitudinal residual two fermion interaction $\Phi$ minus its equivalent, $\mathcal{Q}$, at $\Lambda_s$ (right). Both quantities exhibit very small values, showing that $\mathcal{A}$ and $\Phi$ do not deviate significantly from $\widetilde{\mathcal{Q}}$ and $\mathcal{Q}$, respectively.}
    \label{fig: final vertices}
\end{figure}

\begin{figure}[t]
    \centering
    \includegraphics[width=0.45\textwidth]{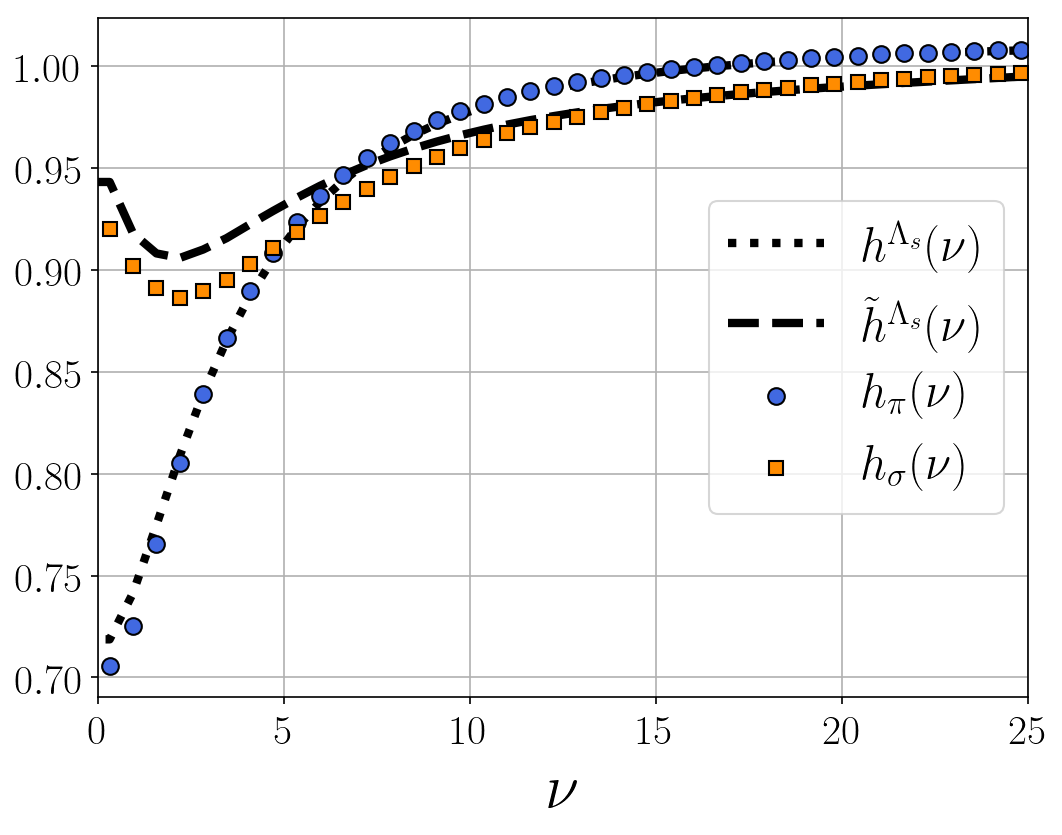}
    \caption{Frequency dependence of Yukawa couplings both at the stopping scale $\Lambda_s$ and in the SSB phase. While $h_\pi$ coincides with $h^{\Lambda_s}$, the longitudinal coupling $h_\sigma$ does not differ significantly from the reduced one at the stopping scale, $\widetilde{h}^{\Lambda_s}$. The continuous lines for $h^{\Lambda_s}$ and $\widetilde{h}^{\Lambda_s}$ are an interpolation through the data calculated on the Matsubara frequencies.}
    \label{fig: hs}
\end{figure}

Within our previously discussed choice of bosonizing $\mathcal{P}^{\Lambda_s}-U$ instead of $\mathcal{P}^{\Lambda_s}$ alone, the self consistent loop introduced in Appendix~\ref{app: loop} converges extremely fast, 15 iterations for example are sufficient to reach a precision of $10^{-7}$ in $\alpha$. 
Once convergence is reached and the gap $\Delta(\nu)$ obtained, we are in the position to evaluate the remaining couplings introduced in Sec.~\ref{sect: bosonic formalism} through their integrated flow equations. In Fig.~\ref{fig: gap} we show the computed frequency dependence of the gap. It interpolates between $\Delta_0=\Delta(\nu\rightarrow 0)$, its value at the Fermi level, and its asymptotic value, that equals the (signed reversed) bare interaction times the \hgl{condensate fraction $\langle\psi_{\downarrow}\psi_{\uparrow}\rangle=$}$\int_\mathbf{k}\langle \psi_{-\mathbf{k},\downarrow}\psi_{\mathbf{k},\uparrow}\rangle$. $\Delta_0$ also represents the gap between the upper and lower Bogoliubov bands. Magnetic and charge fluctuations above the critical scale significantly renormalize the gap with respect to the Hartree-Fock calculation ($\widetilde{V}=-U$ in Eq.~\eqref{eq: gap equation fermionic}), that in the present case coincides with Bardeen-Cooper-Schrieffer (BCS) theory. \hgl{This effect is reminiscent of the Gor'kov-Melik-Barkhudarov correction in weakly coupled superconductors~\cite{Gorkov1961}. The computed frequency dependence of the gap compares qualitatively well with Ref.~\cite{Eberlein2013}, where a more sophisticated truncation of the flow equations has been carried out.}\\
Since $\Delta$ is a spin singlet superfluid gap, and we have chosen $\alpha$ to be real, it obeys 
\begin{equation}
    \Delta(\nu) = \Delta(-\nu) = \Delta^*(-\nu),
\end{equation}
where the first equality comes from the spin singlet nature and the second one from the time reversal symmetry of the effective action. Therefore, the imaginary part of the gap is always zero. By contrast, a magnetic gap would gain, in general, a finite (and antisymmetric in frequency) imaginary part. 
In Fig.~\ref{fig: final vertices} we show the results for the residual two fermion interactions in the longitudinal and transverse channels, together with the total effective interaction in the longitudinal channel, defined as 
\begin{equation}
    V_{\mathcal{A},\nu\nu'}=\frac{h_\sigma(\nu)h_\sigma(\nu')}{m_\sigma}+\mathcal{A}_{\nu\nu'}.
    \label{eq: VA SSB bosonic}
\end{equation}
The analogue of Eq.~\eqref{eq: VA SSB bosonic} for the transverse channel cannot be computed, because the transverse mass $m_\pi$ is zero, in agreement with the Goldstone theorem. The key result is that the residual interactions $\mathcal{A}_{\nu\nu'}$ and $\Phi_{\nu\nu'}$ inherit the frequency structures of $\widetilde{\mathcal{Q}}^{\Lambda_s}_{\nu\nu'}$ and $\mathcal{Q}^{\Lambda_s}_{\nu\nu'}$, respectively, and they are also close to them in values (compare with Fig.~\ref{fig: vertices Lambda s}).
The same occurs for the Yukawa couplings, as shown in Fig.~\ref{fig: hs}. Indeed, the calculated transverse coupling $h_\pi$ does not differ at all from the Yukawa coupling at the stopping scale $h^{\Lambda_s}$. In other words, if instead of solving the self consistent equations, one runs a flow in the SSB phase, the transverse Yukawa coupling will stay the same from $\Lambda_s$ to $\Lambda_\text{fin}$. Furthermore, the longitudinal coupling $h_\sigma$ develops a dependence on the frequency which does not differ significantly from the one of $\widetilde{h}^{\Lambda_s}$.
This feature, at least for our choice of parameters, can lead to some simplifications in the flow equations of Sec.~\ref{sect: bosonic formalism}. Indeed, when running a fRG flow in the SSB phase, one might let flow only the bosonic inverse propagators by keeping the Yukawa couplings and residual interactions fixed at their values, reduced or not, depending on the channel, at the stopping scale. This fact can be crucial to make computational costs lighter when including bosonic fluctuations of the order parameter, which, similarly, do not significantly renormalize Yukawa couplings in the SSB phase~\cite{Obert2013,Bartosch2009}.
\hgl{\subsection{Gap and condensate fraction dependence on the coupling }
In this section, we carry out an analysis of the dependence of the zero-frequency gap $\Delta_0$ on the coupling $U$. In order to obtain a zero temperature estimate, we perform a finite temperature calculation and check that the condition $\Delta_0>T$ is fulfilled. In fact, when this is the case, the superfluid gap is not expected to change significantly by further lowering the temperature, at least within a MF-like calculation. 

In Fig.~\ref{fig: D vs U}, we show the zero-frequency extrapolation of the superfluid gap and the bosonic expectation value $\alpha$ to be compared with the BCS (mean-field) result. The inclusion of magnetic and charge correlations above the stopping scale $\Lambda_s$ renormalizes $\Delta_0$ compared to the BCS result. In particular, as proven by second order perturbation theory in Ref.~\cite{Gorkov1961}, even in the $U\rightarrow 0$ limit the ratio between the ground state gap and its BCS result is expected to be smaller than 1 due to particle-hole fluctuations. Differently, $\alpha$ does not deviate significantly from the mean-field result, as this quantity is not particularly influenced by magnetic and charge fluctuations, but rather by fluctuations of the order parameter, which, in particular at strong coupling, can significantly reduce it~\cite{Bartosch2009}. In the present approach, we include the effect of particle-hole fluctuations and we tackle the frequency dependence of the gap, which are not treated in Refs.~\cite{Bartosch2009,Diehl2007} (which focus on the BEC-BCS crossover in the continuum in 3 dimensions), and~\cite{Obert2013} (2-dimensional lattice model). In these works, however, fluctuations of the order parameter, which are not included in our method, are taken into account. In Ref.~\cite{Eberlein2013}, both particle-hole and order parameter fluctuations, together with the gap frequency dependence, are treated in a rather complicated fRG approach to the 2D attractive Hubbard model, where, however, the Goldstone theorem and the Ward identity turn out to be violated to some extent. We believe our approach to represent a convenient starting point on top of which one can include fluctuations in a systematic manner in order to fulfill the above mentioned fundamental constraints.

\begin{figure}[t]
    \centering
    \includegraphics[width=0.45\textwidth]{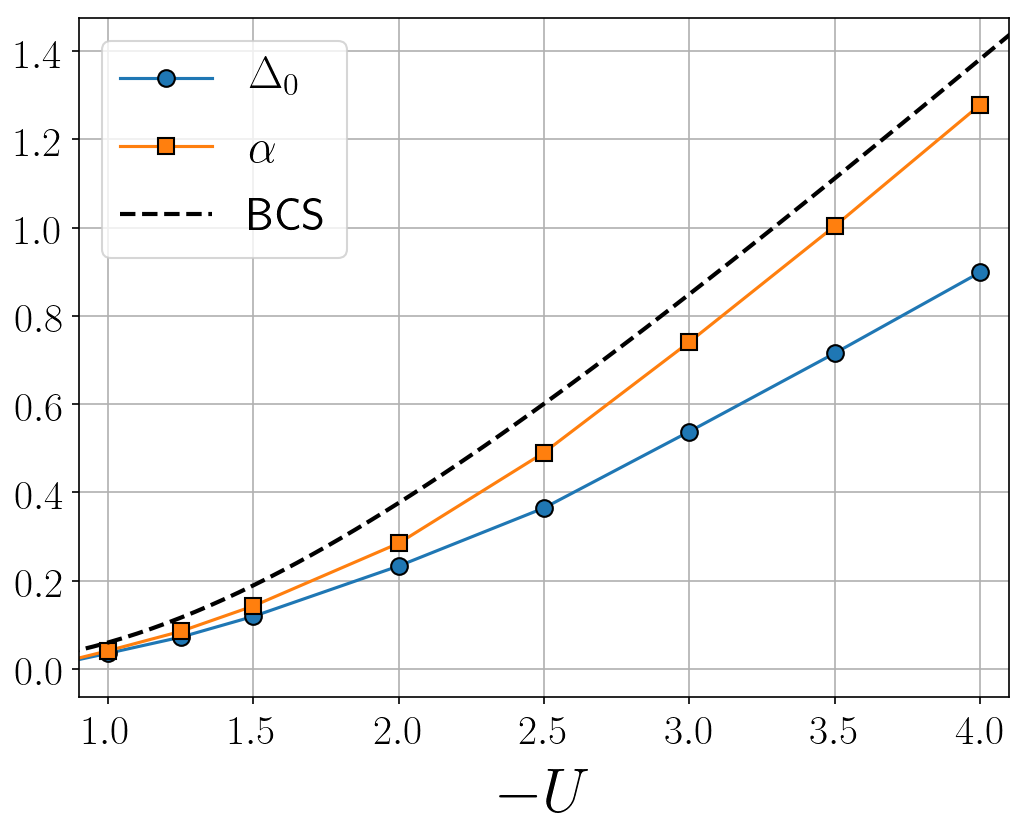}
    \caption{\hgl{Low temperature estimate of the ground state superfluid gap $\Delta_0$ and bosonic expectation value $\alpha$ as a function of the coupling $U$. For $|U|>2t$, the calculations have been performed at $T=0.1t$, while for $t\leq |U|\leq 2t$ we have chosen $T=0.01t$. Results for $|U|<t$ are not shown because the temperature at which $\Delta_0>T$ is fulfilled is hardly reachable by our numerics. The dashed line shows the BCS (for which $\Delta_0=\alpha$) zero temperature result.}}
    \label{fig: D vs U}
\end{figure}
Furthermore, it is interesting to consider the coupling dependence of the condensate fraction $\langle\psi_{\downarrow}\psi_{\uparrow}\rangle$. Within mean-field theory, it evolves from an exponentially small value at weak coupling to $\frac{1}{2}$ at strong coupling, indicating that all the fermions are bound in bosonic pairs which condense. This is an aspect of the well-known paradigm of the BEC-BCS crossover~\cite{Eagles1969,Leggett1980,Nozieres1985}. At half filling, by including quantum fluctuations in the strong coupling regime, it is known that the condensate fraction will be reduced to a value of 0.3, as it has been obtained from the spin wave theory for the Heisenberg model~\cite{Anderson1952}, on which the particle-hole symmetric attractive Hubbard model can be mapped at large $U$.
Within our approach, the condensate fraction is given by
\begin{equation}
    \langle\psi_{\downarrow}\psi_{\uparrow}\rangle=-\lim_{\nu\rightarrow\infty}\frac{\Delta(\nu)}{U}=-\frac{\alpha}{U},
\end{equation}
where in the last line we have used the Ward identity~\eqref{eq: Ward Identity} and the fact that $h_\pi\rightarrow1$ for $\nu\rightarrow\infty$. In Fig.~\ref{fig: QMC comparison} we show the computed condensate fraction and we compare it with the BCS result and with auxiliary field quantum Monte Carlo (AFQMC) data, taken from Ref.~\cite{Karakuzu2018}. At weak coupling ($U=-2t$) we find a good agreement with AFQMC. This is due to the facts that in this regime the order parameter fluctuations, which we do no treat, are weaker, and that the stopping scale $\Lambda_s$ is small and therefore particle-hole fluctuations are better included. At moderate couplings ($U=-3t$, $-4t$) the distance from Monte Carlo data increases due to the increasing strength of fluctuations, the larger stopping scales and the reduction of accuracy of the 1-loop truncation performed in the symmetric phase. 
\begin{figure}[t]
    \centering
    \includegraphics[width=0.45\textwidth]{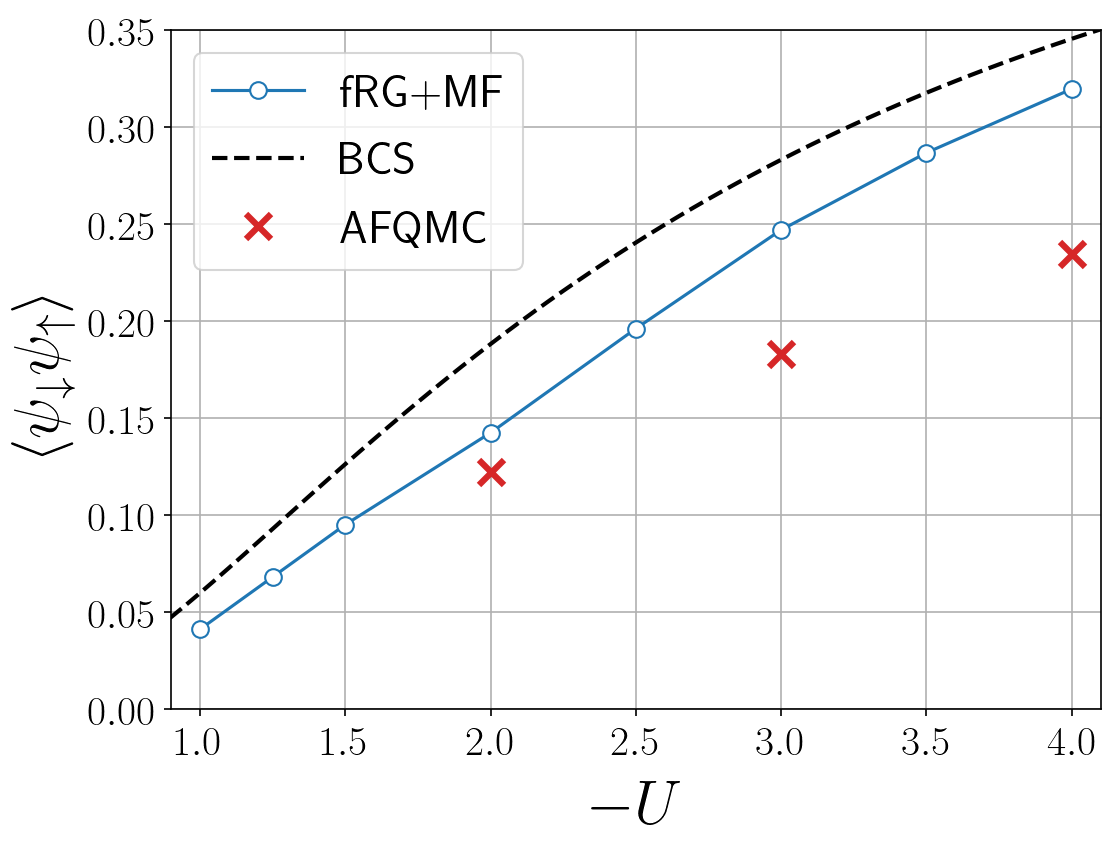}
    \caption{\hgl{Condensate fraction $\langle\psi_{\downarrow}\psi_{\uparrow}\rangle$ \emph{vs} coupling $U$. We indicate the present approach as fRG+MF, and we compare it with BCS theory and AFQMC data, taken from Ref.~\cite{Karakuzu2018}.}}
    \label{fig: QMC comparison}
\end{figure}
}
\section{Conclusion}
\label{sec: conclusion}
We have introduced a truncation of fRG flow equations which, with the introduction of a Hubbard-Stratonovich boson, has been proven to be equivalent to the MF equations obtained in Refs.~\cite{Wang2014,Yamase2016}. These flow equations satisfy fundamental requirements such as the Goldstone theorem and the Ward identities associated with global symmetries, and can be integrated analytically, reducing the calculation of correlation functions in the SSB phase to a couple of self consistent equations for the bosonic expectation value $\alpha$ and the transverse Yukawa coupling $h_\pi$. A necessary step to perform the Hubbard-Stratonovich transformation, on which our method relies, is to extract a factorizable dependence on fermionic variables $k$ and $k'$ from the vertex at the critical scale. A strategy to accomplish this goal has been suggested for a vertex whose dependence on spatial momenta $\mathbf{k}$ and $\mathbf{k}'$ is treated by a form factor expansion, making use of the vertex asymptotics introduced in Ref.~\cite{Wentzell2016}. 
Furthermore, we have tested the feasibility and efficiency of our method on a prototypical model, namely the half-filled attractive Hubbard model in two dimensions, focusing on frequency dependencies of the two fermion interactions, Yukawa couplings and fermionic gap. We have found a good convergence of the iterative scheme proposed. The remaining couplings introduced in our method have been computed after the loop convergence from their integrated flow equations. \hgl{Moreover, we have analyzed the dependence of the gap and of the condensate fraction on the coupling $U$, by comparing our method with previous fRG works and with quantum Monte Carlo data.}

Our method leaves room for applications and extensions. 
First, one can directly apply the MF method, as formulated in this paper, to access the SSB phase in those calculations for which the dependencies on fermionic momenta and/or frequencies cannot be neglected. Some examples are the fRG calculations with a full treatment of fermionic frequencies, within a 1-loop truncation~\cite{Vilardi2017}, in the recent implementations of multiloop fRG~\cite{Tagliavini2019,Hille2020} or in the DMF\textsuperscript{2}RG scheme~\cite{Vilardi2019}. These combinations can be applied to two- or three-dimensional systems. In the former case, even though in 2D order parameter fluctuations are expected to play a decisive role, our method can be useful to get a first, though approximate, picture of the phase diagram. Of particular relevance is the 2D repulsive Hubbard model, used in the context of high-Tc superconductors. An interesting system for the latter case, where bosonic fluctuations are expected to be less relevant, is the 3D attractive Hubbard model, which, thanks to modern techniques, can be experimentally realized in cold atoms setups.  

Secondly, our method constitutes a convenient starting point for the inclusion of bosonic fluctuations of the order parameter, as done for example in Refs.~\cite{Friederich2011,Obert2013}, with the full dependence of the gap, Yukawa couplings and vertices on the fermionic momenta and/or frequencies being kept. In particular, by providing the Hubbard-Stratonovich boson with its own regulator, our MF-truncation of flow equations can be extended to include order parameter fluctuations, which in two spatial dimensions and at finite temperature restore the symmetric phase, in agreement with the Mermin-Wagner theorem. One may also adapt the bosonic field at every fRG step through the \textit{flowing bosonization}~\cite{Friederich2010,Friederich2011}. This can be done by keeping the full frequency dependence of the vertex and Yukawa coupling, by applying the strategy discussed in Sec.~\ref{sec: vertex bosonization} to the flow equation for the vertex. 

Finally, our MF method does not necessarily require a vertex coming from a fRG flow. In particular, one can employ the DMFT vertex, extract the pairing channel~\cite{Rohringer2012} (or any other channel in which symmetry breaking occurs) from it, and apply the same strategy as described in this paper to extract Yukawa and other couplings. This application can be useful to compute those transport quantities~\cite{Bonetti2020} and response functions in the SSB phase which, within the DMFT, require a calculation of vertex corrections~\cite{Georges1996}. The anomalous vertices can be computed also at finite $q$ with a simple generalization of our formulas.
\section*{Acknowledgments}
I am grateful to W. Metzner and D. Vilardi for stimulating discussions and a critical reading of the manuscript.
\appendix
\section{Derivation of flow equations in the bosonic formalism}
\label{app: flow eqs}
In this section we will derive the flow equations used in Sec.~\ref{sect: bosonic formalism}.

We consider only those terms in which the dependence on the center of mass momentum $q$ is fixed to zero by the topology of the relative diagram or that depend only parametrically on it. These diagrams are the only ones necessary to reproduce the MF approximation.

\begin{figure}[t]
    \centering
    \includegraphics[width=0.47\textwidth]{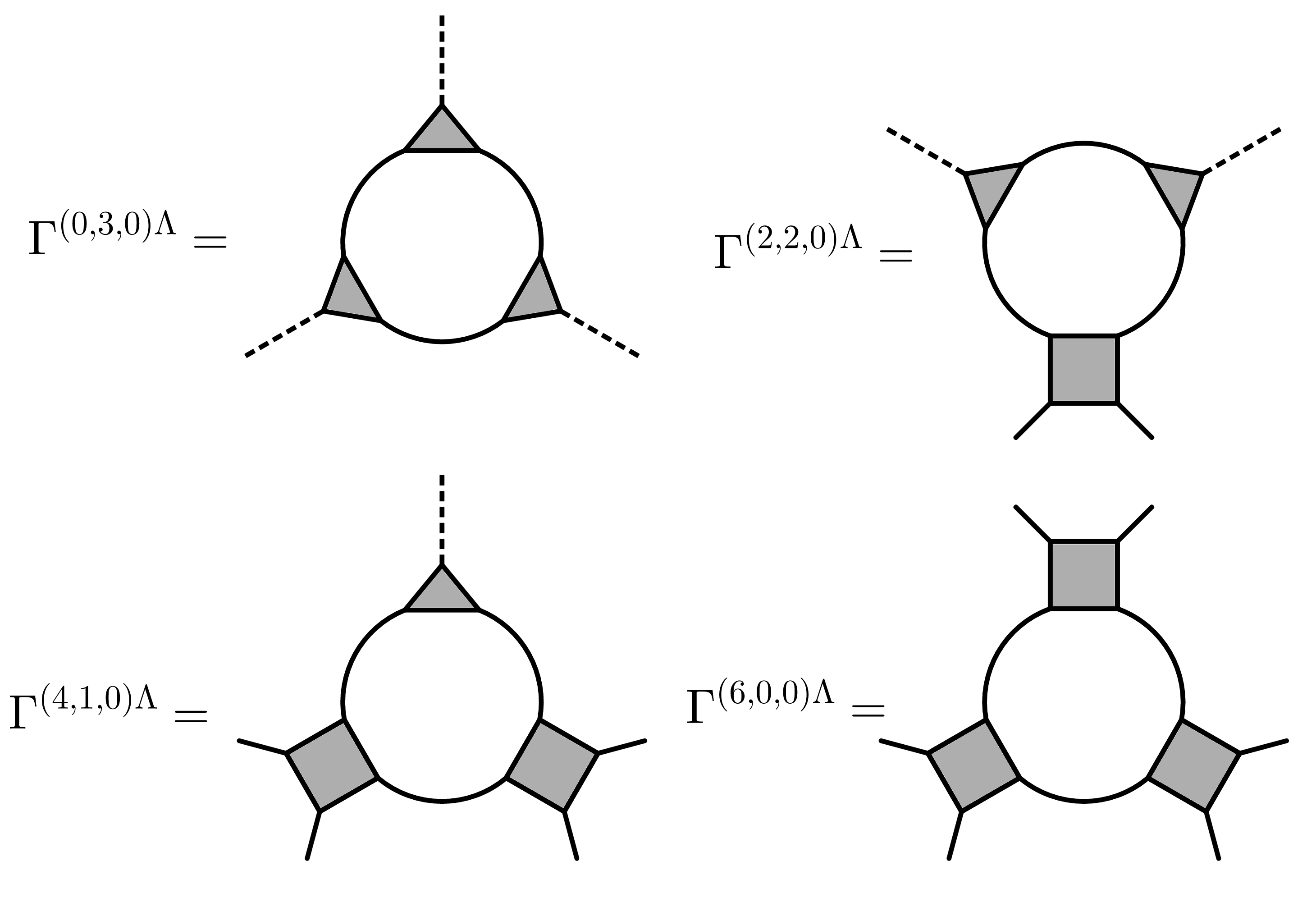}
    \caption{Feynman diagrams describing the Katanin-like approximation higher order correlation functions. The conventions are the same as in Figs.~\ref{fig: flow eqs} and~\ref{fig: flow eqs gaps}.}
    \label{fig: katanin}
\end{figure}

The flow equations will be derived directly from the Wetterich equation~\eqref{eq: Wetterich eq. ferm}, with a slight modification, since we have to keep in mind that the bosonic field $\phi$ acquires a scale dependence due to the scale dependence of its expectation value. The flow equation reads (for real $\alpha^\Lambda$):
\begin{equation}
    \partial_\Lambda\Gamma^\Lambda=\frac{1}{2}\widetilde{\partial}_\Lambda\text{Str}\ln\left[\mathbf{\Gamma}^{(2)\Lambda}+R^\Lambda\right]+\frac{\delta\Gamma^\Lambda}{\delta\sigma_{q=0}} \, \partial_\Lambda \alpha^\Lambda,
    \label{eq: Wetterich eq.}
\end{equation}
where $\mathbf{\Gamma}^{(2)\Lambda}$ is the matrix of the second derivatives of the action with respect to the fields, the supertrace Str includes a minus sign when tracing over fermionic variables. The first equation we derive is the one for the flowing expectation value $\alpha^\Lambda$. This is obtained by requiring that the one-point function for $\sigma_q$ vanishes. Taking the $\sigma_q$ derivative in Eq.~\eqref{eq: Wetterich eq.} and setting the fields to zero, we have
\begin{equation}
    \begin{split}
        &\partial_\Lambda\Gamma^{(0,1,0)\Lambda}(q=0)\equiv\partial_\Lambda\frac{\delta\Gamma^\Lambda}{\delta\sigma_{q=0}}\bigg\lvert_{\Psi,\overline{\Psi},\sigma,\pi=0}\\
        &=-\int_k h^\Lambda_\sigma(k;0)\, \widetilde{\partial}_\Lambda F^\Lambda(k)+m^\Lambda_\sigma(0)\,\partial_\Lambda\alpha^\Lambda=0,
    \end{split}
    \label{eq: alpha eq app}
\end{equation}
where we have defined
\begin{equation}
    \Gamma^{(2n_1,n_2,n_3)\Lambda}=\frac{\delta^{(2n_1+n_2+n_3)}\Gamma^\Lambda}{\left(\delta\overline{\Psi}\right)^{n_1}\left(\delta\Psi\right)^{n_1}\left(\delta\sigma\right)^{n_2}\left(\delta\pi\right)^{n_3}}.    
\end{equation}
From Eq.~\eqref{eq: alpha eq app} we get the flow equation for $\alpha^\Lambda$.
\begin{equation}
    \partial_\Lambda\alpha^\Lambda=\frac{1}{m^\Lambda_\sigma(0)}\int_k h^\Lambda_\sigma(k;0)\, \widetilde{\partial}_\Lambda F^\Lambda(k).
    \label{eq: alpha flow}
\end{equation}
The MF flow equation for the fermionic gap reads
\begin{equation}
    \begin{split}
        \partial_\Lambda \Delta^\Lambda(k)&= 
        \int_{k'}\mathcal{A}^\Lambda(k,k';0)\,\widetilde{\partial}_\Lambda F^\Lambda(k')\\
        &+\partial_\Lambda\alpha^\Lambda\,h_\sigma^\Lambda(k;0),    
    \end{split}
    \label{eq: gap equation bosonic}
\end{equation}
with $\mathcal{A}^\Lambda$ being the residual two fermion interaction in the longitudinal channel. 
The equation for the inverse propagator of the $\sigma_q$ boson is
\begin{equation}
    \begin{split}
        \partial_\Lambda m_\sigma^\Lambda(q)=&\int_p h_\sigma^\Lambda(p;q)\left[\widetilde{\partial}_\Lambda\Pi^\Lambda_{11}(p;q)\right] h_\sigma^\Lambda(p;q)\\
        +&\int_p \Gamma^{(2,2,0)\Lambda}(p,0,q)\,\widetilde{\partial}_\Lambda F^\Lambda(p)\\
        +&\partial_\Lambda\alpha^\Lambda\,\Gamma^{(0,3,0)\Lambda}(q,0),
        \label{eq: full P sigma flow}
    \end{split}
\end{equation}
where we have defined the bubble at finite momentum $q$ as
\begin{equation}
    \Pi^\Lambda_{\alpha\beta}(k;q)=-\frac{1}{2}\Tr\left[\tau^\alpha\mathbf{G}^\Lambda(k)\tau^\beta\mathbf{G}^\Lambda(k-q)\right], 
\end{equation}
$\Gamma^{(0,3,0)\Lambda}$ is an interaction among three $\sigma$ bosons and $\Gamma^{(2,2,0)\Lambda}$ couples one fermion and 2 longitudinal bosonic fluctuations. 
The equation for the longitudinal Yukawa coupling is 
\begin{equation}
    \begin{split}
        \partial_\Lambda h^\Lambda_\sigma(k;q)=&\int_p\mathcal{A}^\Lambda(k,p;q)\left[\widetilde{\partial}_\Lambda\Pi^\Lambda_{11}(p;q)\right]h^\Lambda_\sigma(p,q)\\
        +&\int_{k'}\Gamma^{(4,1,0)\Lambda}(k,p,q,0)\,\widetilde{\partial}_\Lambda F^\Lambda(p)\\
        +&\partial_\Lambda\alpha^\Lambda\,\Gamma^{(2,2,0)\Lambda}(k,q,0),
        \label{eq: full h_sigma flow}
    \end{split}
\end{equation}
where $\Gamma^{(4,1,0)\Lambda}$ is a coupling among 2 fermions and one $\sigma$ boson.
The flow equation for the coupling $\mathcal{A}^\Lambda$ reads instead 
\begin{equation}
    \begin{split}
        \partial_\Lambda\mathcal{A}^\Lambda(k,k';q)=&\int_p\mathcal{A}^\Lambda(k,p;q)\left[\widetilde{\partial}_\Lambda\Pi^\Lambda_{11}(p;q)\right]\mathcal{A}^\Lambda(p,k';q)\\
        +&\int_{p}\Gamma^{(6,0,0)\Lambda}(k,k',q,p,0)\,\widetilde{\partial}_\Lambda F^\Lambda(p)\\
        +&\partial_\Lambda\alpha^\Lambda\,\Gamma^{(4,1,0)\Lambda}(k,k',q,q),
        \label{eq: full A flow}
    \end{split}
\end{equation}
with $\Gamma^{(6,0,0)\Lambda}$ the 3-fermion coupling. We recall that in all the above flow equations, we have considered only the terms in which the center of mass momentum $q$ enters parametrically in the equations. This means that we have assigned to the flow equation for $\mathcal{A}^\Lambda$ only contributions in the particle-particle channel and we have neglected in all flow equations all the terms that contain a loop with the normal single scale propagator $\widetilde{\partial}_\Lambda G^\Lambda(k)$. Within a reduced model, where the bare interaction is nonzero only for $q=0$ scattering processes, the mean-field is the exact solution and one can prove that, due to the reduced phase space, only the diagrams that we have considered in our truncation of the flow equations survive~\cite{Salmhofer2004}. In order to treat the higher order couplings, $\Gamma^{(0,3,0)\Lambda}$, $\Gamma^{(2,2,0)\Lambda}$, $\Gamma^{(4,1,0)\Lambda}$  and $\Gamma^{(6,0,0)\Lambda}$, one can approximate their flow equations in order to make them integrable in way similar to Katanin's approximation for the 3-fermion coupling. The integrated results are the fermionic loop integrals schematically shown in Fig.~\ref{fig: katanin}. Skipping any calculation, we just state that this approximation allows for absorbing the second and third terms on the right hand-side of Eqs.~\eqref{eq: full P sigma flow},~\eqref{eq: full h_sigma flow} and~\eqref{eq: full A flow} into the first one just by replacing $\widetilde{\partial}_\Lambda\Pi^\Lambda_{11}$ with its full derivative $\partial_\Lambda\Pi^\Lambda_{11}$. In summary:
\begin{align}
    &\partial_\Lambda m_\sigma^\Lambda(q)=\int_p h_\sigma^\Lambda(p;q)\left[\partial_\Lambda\Pi^\Lambda_{11}(p;q)\right] h_\sigma^\Lambda(p;q),\\
    &\partial_\Lambda h^\Lambda_\sigma(k;q)=\int_p\mathcal{A}^\Lambda(k,p;q)\left[\partial_\Lambda\Pi^\Lambda_{11}(p;q)\right]h^\Lambda_\sigma(p;q),\\
    &\partial_\Lambda\mathcal{A}^\Lambda(k,k';q)=\int_p\mathcal{A}^\Lambda(k,p;q)\left[\partial_\Lambda\Pi^\Lambda_{11}(p;q)\right]\mathcal{A}^\Lambda(p,k';q).
\end{align}
With a similar approach, one can derive the flow equations for the transverse couplings:
\begin{align}
        &\partial_\Lambda m_\pi^\Lambda(q)=\int_p h_\pi^\Lambda(p;q)\left[\partial_\Lambda\Pi^\Lambda_{22}(p;q)\right] h_\pi^\Lambda(p;q),\\
        &\partial_\Lambda h^\Lambda_\pi(k;q)=\int_p\Phi^\Lambda(k,p;q)\left[\partial_\Lambda\Pi^\Lambda_{22}(p;q)\right]h_\pi^\Lambda(p;q),\\
        &\partial_\Lambda\Phi^\Lambda(k,k';q)=\int_p\Phi^\Lambda(k,p;q)\left[\partial_\Lambda\Pi^\Lambda_{22}(p;q)\right]\Phi^\Lambda(p,k';q).
\end{align}
\section{Calculation of the irreducible vertex in the bosonic formalism}
\label{app: irr V bosonic formalism}
In this appendix we provide a proof of Eq.~\eqref{eq: irr V bosonic formalism} by making use of matrix notation. 
If the full vertex can be decomposed as in Eq.~\eqref{eq: vertex at Lambda crit}
\begin{equation}
    V=\mathcal{Q}+\frac{h [h]^T}{m},
\end{equation}
we can plug this relation into the definition of the irreducible vertex, Eq.~\eqref{eq: irr vertex fermionic}. With some algebra we obtain
\begin{equation}
    \begin{split}
        \widetilde{V}=&\left[1+V\Pi\right]^{-1}V=\\
        =&\left[1+\frac{\widetilde{h} [h]^T}{m}\Pi\right]^{-1}\left[\widetilde{\mathcal{Q}}+\frac{\widetilde{h} [h]^T}{m}\right],
    \end{split}
    \label{eq: V tilde appendix I}
\end{equation}
where in the last line we have inserted a representation of the identity,
\begin{equation}
    1=\left[1+\mathcal{Q}\Pi\right]\left[1+\mathcal{Q}\Pi\right]^{-1},  
\end{equation}
in between the two matrices and we have made use of definitions~\eqref{eq: reduced C tilde} and~\eqref{eq: reduced yukawa}. With a bit of simple algebra, we can analytically invert the matrix on the left in the last line of Eq.~\eqref{eq: V tilde appendix I}, obtaining
\begin{equation}
    \left[1+\frac{\widetilde{h} [h]^T}{m}\Pi\right]^{-1}=1-\frac{\widetilde{h} [h]^T}{\widetilde{m}}\Pi,
\end{equation}
%
Where $\widetilde{m}$ is defined in Eq.~\eqref{eq: reduced mass P tilde}. By plugging this result into Eq.~\eqref{eq: V tilde appendix I}, we finally obtain
\begin{equation}
    \widetilde{V}=\widetilde{\mathcal{Q}}+\frac{\widetilde{h} [\widetilde{h}]^T}{\widetilde{m}},
\end{equation}
that is the result of Eq.~\eqref{eq: irr V bosonic formalism}.\\
\vskip 1cm
\section{Algorithm for the calculation of the superfluid gap}
\label{app: loop}
The formalism described in Sec.~\ref{sect: bosonic formalism} allows us to formulate a minimal set of closed equations required for the calculation of the gap. We drop the $\Lambda$ superscript, assuming that we have reached the final scale. The gap can be computed using the Ward identity, so we can reduce ourselves to a single self consistent equation for $\alpha$, that is a single scalar quantity, and another one for $h_\pi$, momentum dependent. The equation for $\alpha$ is Eq.~\eqref{eq: alpha solution}. The transverse Yukawa coupling is calculated through Eq.~\eqref{eq: h_pi}. The equations are coupled since the superfluid gap $\Delta=\alpha h_\pi$ appears in the r.h.s of both.\\
We propose an iterative loop to solve the above mentioned equations. By starting with the initial conditions $\alpha^{(0)}=0$ and $h_\pi^{(0)}(k)=0$, we update the transverse Yukawa coupling at every loop iteration $i$ according to Eq.~\eqref{eq: h_pi}, that can be reformulated in the following algorithmic form:
\begin{equation}
    h_\pi^{(i+1)}(k)=\int_{k'} \left[M^{(i)}(k,k')\right]^{-1}\,\widetilde{h}^{\Lambda_s}(k'), 
    \label{eq: h_pi loop}
\end{equation}
with the matrix $M^{(i)}$ defined as 
\begin{equation}
    M^{(i)}(k,k')=\delta_{k,k'}-\mathcal{\widetilde{Q}}^{\Lambda_s}(k,k')\,\Pi_{22}^{(i)}(k';\alpha^{(i)}),
\end{equation}
and the $22$-bubble rewritten as
\begin{equation}
    \Pi_{22}^{(i)}(k;\alpha) = \frac{1}{G^{-1}(k)G^{-1}(-k)+\alpha^2\left[h_\pi^{(i)}(k)\right]^2},
\end{equation}
with $G(k)$ defined in Eq.~\eqref{eq: G at Lambda_s}.
Eq.~\eqref{eq: h_pi loop} is not solved self consistently at every loop iteration $i$, because we have chosen to evaluate the r.h.s with $h_\pi$ at the previous iteration. $\alpha^{(i+1)}$ is calculated by self consistently solving
\begin{equation}
    1=\frac{1}{\widetilde{m}^{\Lambda_s}}\int_k \widetilde{h}^{\Lambda_s}(k)\, \Pi_{22}^{(i+1)}(k;\alpha) \,h_\pi^{(i+1)}(k)
    \label{eq: alpha loop}
\end{equation}
for $\alpha$. The equation above is nothing but Eq.~\eqref{eq: alpha solution} where the solution $\alpha=0$ has been factorized away.
The loop consisting of Eqs.~\eqref{eq: h_pi loop} and~\eqref{eq: alpha loop} must be repeated until convergence is reached in $\alpha$ and, subsequently, in $h_\pi$.
This formulation of self consistent equations is not computationally lighter than the one in the fermionic formalism, but more easily controllable, as one can split the frequency and momentum dependence of the gap (through $h_\pi$) from the strength of the order ($\alpha$). Moreover, thanks to the fact that $h_\pi$ is updated with an explicit expression, namely Eq.~\eqref{eq: h_pi loop}, that is in general a well behaved function of $k$, the frequency and momentum dependence of the gap is assured to be under control.
\section{Flow equations in the symmetric phase}
\label{app: flow eqs symm phase}
The flow equations for the three channels, $\mathcal{C}^\Lambda$, $\mathcal{M}^\Lambda$ and $\mathcal{P}^\Lambda$ in the symmetric phase are similar to those obtained in Ref.~\cite{Vilardi2017} for the repulsive Hubbard model. For the charge channel we have 
\begin{widetext}
\begin{equation}
    \begin{split}
        \partial_\Lambda\mathcal{C}^\Lambda_{\nu\nu'}(\mathbf{q},\Omega)=-\sum_{\omega}L^{c,\Lambda}_{\nu,\omega+\Omega}(\mathbf{q},\Omega)\left[\widetilde{\partial}_\Lambda\chi^{c,\Lambda}_{\omega}(\mathbf{q},\Omega)\right]L^{c,\Lambda}_{\omega\nu'}(\mathbf{q},\Omega),
    \end{split}
\end{equation}
where 
\begin{equation}
    \begin{split}
        L^{c,\Lambda}_{\nu\nu'}(\mathbf{q},\Omega)= U - \mathcal{C}^\Lambda_{\nu\nu'}(\mathbf{q},\Omega) 
        +\scaleint{8ex}_{\mkern-15mu \mathbf{p}}\Bigg[\frac{1}{2}
        \mathcal{C}^\Lambda_{\nu\nu'}(\mathbf{p},\nu'-\nu-\Omega)
        +\frac{3}{2}\mathcal{M}^\Lambda_{\nu\nu'}(\mathbf{p},\nu'-\nu-\Omega)
        -\mathcal{P}^\Lambda_{\nu,\nu+\Omega}(\mathbf{p},\nu+\nu')\Bigg],
    \end{split}
\end{equation}
and 
\begin{equation}
    \chi^{c,\Lambda}_{\nu}(\mathbf{q},\Omega)=T\int_\mathbf{k}G_0^\Lambda\left(\mathbf{k}+\mathbf{q},\nu+\Omega\right)G_0^\Lambda\left(\mathbf{k},\nu\right),
\end{equation}
with $G_0^\Lambda(k)=\frac{1}{Q_0^\Lambda(k)}=\frac{\nu^2}{\nu^2+\Lambda^2}\frac{1}{i\nu-\xi_\mathbf{k}}$, and $\int_\mathbf{k}=\int \frac{d^2\mathbf{k}}{(2\pi)^2}$.
Similarly, the flow equation for the magnetic channel is given by
\begin{equation}
    \begin{split}
        \partial_\Lambda\mathcal{M}^\Lambda_{\nu\nu'}(\mathbf{q},\Omega)=-\sum_{\omega}L^{m,\Lambda}_{\nu,\omega+\Omega}(\mathbf{q},\Omega)\left[\widetilde{\partial}_\Lambda\chi^{m,\Lambda}_{\omega}(\mathbf{q},\Omega)\right]L^{m,\Lambda}_{\omega\nu'}(\mathbf{q},\Omega),
    \end{split}
\end{equation}
with $\chi^{m,\Lambda}_{\omega}(\mathbf{q},\Omega)=\chi^{c,\Lambda}_{\omega}(\mathbf{q},\Omega)$, and
\begin{equation}
    \begin{split}
        L^{m,\Lambda}_{\nu\nu'}(\mathbf{q},\Omega)= -U - \mathcal{M}^\Lambda_{\nu\nu'}(\mathbf{q},\Omega) + \scaleint{8ex}_{\mkern-15mu \mathbf{p}}
        \Bigg[\frac{1}{2}
        \mathcal{C}^\Lambda_{\nu\nu'}(\mathbf{p},\nu'-\nu-\Omega)
        -\frac{1}{2}\mathcal{M}^\Lambda_{\nu\nu'}(\mathbf{p},\nu'-\nu-\Omega)
        +\mathcal{P}^\Lambda_{\nu,\nu+\Omega}(\mathbf{p},\nu+\nu')\Bigg].
    \end{split}
\end{equation}
Finally, the flow equation for the pairing channel reads
\begin{equation}
    \begin{split}
        \partial_\Lambda\mathcal{P}^\Lambda_{\nu\nu'}(\mathbf{q},\Omega)=\sum_{\omega}L^{p,\Lambda}_{\nu\omega}(\mathbf{q},\Omega)\left[\widetilde{\partial}_\Lambda\chi^{p,\Lambda}_{\omega}(\mathbf{q},\Omega)\right]L^{p,\Lambda}_{\omega\nu'}(\mathbf{q},\Omega),
    \end{split}
\end{equation}
where we have projected onto the singlet component of the pairing, that is
\begin{equation}
    L^{p,\Lambda}_{\nu\nu'}(\mathbf{q},\Omega)=\frac{\overline{L}^{p,\Lambda}_{\nu\nu'}(\mathbf{q},\Omega)+\overline{L}^{p,\Lambda}_{\nu,\Omega-\nu'}(\mathbf{q},\Omega)}{2}, 
\end{equation}
with 
\begin{equation}
    \begin{split}
        \overline{L}^{p,\Lambda}_{\nu\nu'}(\mathbf{q},\Omega)= -U + \mathcal{P}^\Lambda_{\nu\nu'}(\mathbf{q},\Omega) 
        + \scaleint{8ex}_{\mkern-15mu \mathbf{p}}
        \Bigg[\frac{1}{2}
        \mathcal{C}^\Lambda_{\nu,\Omega-\nu}(\mathbf{p},\nu'-\nu)
        -\frac{1}{2}\mathcal{M}^\Lambda_{\nu,\Omega-\nu}(\mathbf{p},\nu'-\nu)
        -\mathcal{M}^\Lambda_{\nu,\Omega-\nu}(\mathbf{p},\Omega-\nu-\nu')\Bigg],
    \end{split}
\end{equation}
and
\begin{equation}
    \chi^{p,\Lambda}_{\nu}(\mathbf{q},\Omega)=T\int_\mathbf{k}G_0^\Lambda\left(\mathbf{k},\nu\right)G_0^\Lambda\left(\mathbf{q}-\mathbf{k},\Omega-\nu\right).
\end{equation}
\end{widetext}
\bibliography{main.bib}
\end{document}